\documentclass[11pt,twoside]{jcap}
\usepackage{epsfig}
\usepackage{amssymb}
\usepackage{fancyhed}

\newcommand{\ftn}{\footnotesize}
\newcommand{\nsz}{\normalsize}
\newcommand{\ssz}{\scriptsize}

\newcommand{\vH}{{\mbox{$\bar H$}}}
\newcommand{\vHi}{{\mbox{$\bar H_{_{\rm I}}$}}}
\newcommand{\vrho}{{\mbox{$\bar\rho$}}}
\newcommand{\vq}{{\mbox{$\bar q$}}}
\newcommand{\vQ}{{\mbox{$\bar Q$}}}
\newcommand{\vV}{{\mbox{$\bar V$}}}
\newcommand{\vVo}{{\mbox{$\bar V_o$}}}

\newcommand{\vm}{{\mbox{$\bar m$}}}
\newcommand{\vti}{{\mbox{$\vtau_{_{\rm I}}$}}}
\newcommand{\vtif}{{\mbox{$\vtauf_{_{\rm I}}$}}}
\newcommand{\vtf}{{\mbox{$\vtau_{_{\rm F}}$}}}
\newcommand{\vtns}{{\mbox{$\vtau_{_{\rm NS}}$}}}
\newcommand{\vtkr}{{\mbox{$\vtau_{_{\rm KR}}$}}}
\newcommand{\vtb}{{\mbox{$\vtau_{_{\rm B}}$}}}

\newcommand{\vsv}{{\mbox{ $\overline{\langle \sigma v \rangle}$}}}
\newcommand{\sv}{{\mbox{$\langle \sigma v \rangle$}}}

\def\openep{\leavevmode\hbox{\normalsize$\iota$\kern-3.8pt$^$-}}
\def\vtau{\leavevmode\hbox{\normalsize$\tau$\kern-5.pt$\iota$}}
\def\vtauf{\leavevmode\hbox{\ftn$\tau$\kern-4.pt$\iota$}}

\def\beq{\begin{equation}}
\def\eeq{\end{equation}}
\def\bea{\begin{eqnarray}}
\def\eea{\end{eqnarray}}

\begin{document}

\title{\huge Q{\Large UINTESSENTIAL} K{\Large INATION  AND} \\\vspace{1pt}
C{\Large OLD}  D\kern-1.pt{\Large ARK}
M\kern-1.pt{\Large ATTER} A\Large BUNDANCE}

\author{\Large C. P\kern-1.5pt\nsz ALLIS}

\address{Physics Division, School of Technology,\\
Aristotle University of Thessaloniki, \\
541 24 Thessaloniki, GREECE}
\ead{kpallis@auth.gr}


\begin{abstract}{\vspace{5pt}\par The generation of a kination-dominated
phase by a quintessential exponen-tial model is investigated and
the parameters of the model are restricted so that a number of
observational constraints (originating from nucleosynthesis, the
present acceleration of the universe and the dark-energy-density
parameter) are satisfied. The decoupling of a thermal cold dark
matter particle during the period of kination is analyzed, the
relic density is calculated both numerically and semi-analytically
and the results are compared with each other. It is argued that
the enhancement, with respect to the standard paradigm, of the
cold dark matter abundance can be expressed as a function of the
quintessential density parameter at the onset of nucleosynthesis.
We find that values of the latter quantity close to its upper
bound require the thermal-averaged cross section times the
velocity of the cold relic to be almost three orders of magnitude
larger than this needed in the standard scenario so as
compatibility with the cold dark matter constraint is achieved.}
\end{abstract}

\keyw{Cosmology, Dark Energy, Dark Matter} 

\publishedin{{\sl J. Cosmol. Astropart. Phys.} {\bf 015}, 10
(2005)}

\maketitle

\tableofcontents \vskip0.10cm\noindent\rule\textwidth{.4pt}

\setcounter{page}{1} \pagestyle{fancyplain}


\rhead[\fancyplain{}{ \bf \thepage}]{\fancyplain{}{Q{\ftn
UINTESSENTIAL} K{\ftn INATION AND} CDM A{\ftn BUNDANCE}}}
\lhead[\fancyplain{}{ \leftmark}]{\fancyplain{}{\bf \thepage}}
\cfoot{}

\section{I{\ftn NTRODUCTION}}\label{intro}

\hspace{.67cm} A plethora of recent data \cite{wmap, snae} has
indicated that the two major components of the present universe
are the Cold (mainly \cite{wmapl}) Dark Matter (CDM) and the Dark
Energy (DE) with density parameters \cite{wmap}, respectively:
\beq {\sf (a)}~~\Omega_{\rm CDM}=0.24\pm0.1~~\mbox{and}~~{\sf
(b)}~~\Omega_{\rm DE}=0.73\pm0.12, \label{cdmba}\eeq
at $95\%$ confidence level (c.l.). The identification of these two
unknown substances consists one of the most tantalizing enigmas of
the modern cosmo-particle theories.

As regards CDM, the most natural candidates \cite{candidates} are
the weekly interacting massive particles, $\chi$'s. The most
popular of these is the lightest supersymmetric (SUSY) particle
(LSP) \cite{goldberg}. However, the extra dimensional (ED)
theories give rise to new CDM candidates \cite{lkk, lzk, branon}.
According to the standard cosmological scenario (SC) \cite{kolb},
$\chi$'s (i) are produced through thermal scatterings in the
plasma, (ii) reach chemical equilibrium with plasma and (iii)
decouple from the cosmic fluid  during the radiation-dominated
(RD) era (note that these assumptions are, also, naturally valid
in the case of the so-called second Randall-Sundrum \cite{extra}
model, provided that the brane-tension is constrained to rather
high values \cite{seto}). The viability of other CDM candidates
(like axions \cite{axion}, axino \cite{axino}, gravitino
\cite{gravitino}, quintessino \cite{qino}) requires a somehow
different cosmological set-up, which we do not consider in our
analysis. In light of eq.~(\ref{cdmba}), the $\chi$-relic density,
$\Omega_{\chi}h_0^2$, is to satisfy the following range of values:
\beq {\sf (a)}~~0.09\lesssim \Omega_{\chi}h_0^2~~\mbox{and}~~ {\sf
(b)}~~\Omega_{\chi}h_0^2\lesssim0.13. \label{cdmb}\eeq

As regards DE, quintessence \cite{early}, a slowly evolving scalar
field, has recently attracted much attention (for reviews, see
ref.~\cite{der}). The scalar field is supposed to roll down its
potential undergoing three phases during its cosmological
evolution: Initially its kinetic energy, which decreases faster
than the radiation, dominates and gives rise to a possible novel
period in the universal history termed ``kination''
\cite{kination}. Then, the scalar field freezes to a value close
to Planck scale and by now its potential energy, adjusted so that
eq.~(\ref{cdmba}{\sf b}) is met, becomes dominant. Such an
adjustment, which certainly does not resolve satisfactorily the
coincidence problem, is unavoidable in quintessential models (for
related suggestions, see refs.~\cite{vmp, cp}). Other shortcomings
such as the lightness of the scalar field \cite{st} or the time
variation of the gauge coupling constants \cite{carroll} are
currently under investigation.

Be that as it may, the viability of a quintessential scenario can
be controlled by imposing some observational constraints
\cite{brazil}, arising from nucleosynthesis, acceleration of the
universe and the DE density parameter. Unfortunately no
full-satisfactory potential exists, to date (for comparative
explorations of various potentials, see refs.~\cite{silogi,
german}). E.g., the inverse power potential \cite{inv} although
provides a tracker-type solution \cite{attr} does not fit well
\cite{brax} the present-day value \cite{wmap, snae} of the
quintessence-equation-of-state parameter. Phenomenologically more
robust \cite{german} is the supergravity-inspired \cite{brax,
expo} potential without, however, to allow a zero minimum of the
potential \cite{brax, wmapl}. Also, in both cases, the generation
in the early universe of a kination-dominated (KD) expansion
consistent with the fulfillment of the requirements above is
rather questionable \cite{salati, rosati}. For these reasons, we
decide to examine the simplest exponential potential \cite{wet,
expo}, which, although does not possess a tracker-type solution
\cite{attr, salati}, it can produce a viable present-day cosmology
in conjunction with the domination of an early KD era, for a
reasonable region of initial conditions
\cite{brazil,german,cline}.

The departure from the SC, caused by the implementation of a
quintessential KD epoch can modify the $\Omega_{\chi}h_0^2$
calculation, which (as, already, emphasized \cite{Kam,fornengo,
snr, masiero}) crucially dependents on the adopted assumptions. If
the quintessential KD phase dominates over the radiation (a
condition indispensable for the quintessential inflationary
model-building \cite{qinf,qinfHI, dimopoulos}), the assumption
(iii) of the SC is lifted (note that the assumptions (i) and (ii)
are maintained). As a consequence, an increase to
$\Omega_{\chi}h_0^2$ with respect to (w.r.t) its value in the SC
is implied. This phenomenon was first pointed out in
ref.~\cite{salati} and was explored in ref.~\cite{prof} for the
parameters of the exponential potential, which support a global
attractor \cite{ferreira}. There \cite{prof}, $\Omega_{\chi}h_0^2$
was calculated numerically for a couple of SUSY models which
resurrect higgsino \cite{su5} or wino \cite{wells} LSP and can
yield acceptable $\Omega_{\chi}h_0^2$.

Contrary to ref.~\cite{prof}, we focus on the range of the
exponential-potential parameters, which ensures a late-time
attractor together with an early KD regime (see sec.~\ref{dynq})
and can lead to a simultaneous satisfaction of several
observational data (see secs.~\ref{reqq} and \ref{Qpar}). We then,
present a ``unified'' (using the same independent variable)
description of the cosmological evolution of the quintessence
field and the $\chi$ decoupling (see secs.~\ref{dynq} and
\ref{Seqs}). The relevant equations are solved both numerically
(see secs.~\ref{Beqs} and \ref{Neqs}) and semi-analytically (see
secs.~\ref{dynq} and \ref{Seqs}) and the results are compared with
each other (see secs.~\ref{Qev} and \ref{omenh}). Finally, we
demonstrate the crucial correlation between the
$\Omega_{\chi}h_0^2$ enhancement (w.r.t the one in the SC) and the
quintessential density parameter at the eve of nucleosynthesis
(see sec.~\ref{omenh}) and we restrict the parameters imposing all
the DE and CDM constraints (see sec.~\ref{NTR}) without, however,
to adopt a specific particle model. We showed that values of the
quintessential density parameter at the former point close to its
upper bound require the thermal-averaged cross section times the
velocity of $\chi$ to be almost three orders of magnitude larger
than this needed in the SC.

The framework of the quintessential cosmology is described in
sec.~\ref{cosmoq}, while our numerical and semi-analytical
$\Omega_{\chi}h_0^2$ calculations are displayed in
sec.~\ref{cdmhr}. Some numerical applications are presented in
sec.~\ref{ap}. Finally, sec.~\ref{con} summarizes our results and
discusses some open questions. Throughout the text and the
formulas, brackets are used by applying disjunctive
correspondence, natural units ($\hbar=c=k_{\rm B}=1$) are assumed,
the subscript or superscript $0$ is referred to present-day values
and $\ln~[\log]$ stands for logarithm with basis $e~[10]$.

\section{Q{\ftn UINTESSENTIAL} C{\ftn OSMOLOGY}}
\label{cosmoq} \setcounter{equation}{0}

\hspace{.67cm} We briefly describe the equations which govern the
evolution of the universe in the presence of quintessence
(sec.~\ref{Beqs}), the phases which the quintessence field
undergoes during its evolution (sec.~\ref{dynq}) and the
requirements which a successful quintessential scenario is to
satisfy (sec.~\ref{reqq}).

\subsection{R{\ssz ELEVANT} E{\ssz QUATIONS}} \label{Beqs}

\hspace{.67cm} According to the quintessential scenario, we assume
the existence of a spatially homogeneous, scalar field $q$ (not to
be confused with the deceleration parameter \cite{wmapl}) which
obeys the Klein-Gordon equation. We below present its archetypal
form and then we derive simplified forms which facilitate its
numerical integration. Finally, we specify the used initial
conditions and we define the useful extracted quantities.
\subsubsection{Initial form.} The homogeneous Klein-Gordon
equation in a cosmological set-up is
\beq \mbox{\sf (a)}~~\ddot q+3H\dot
q+V_{,q}=0,~~\mbox{where}~~\mbox{\sf (b)}~~V=V_o e^{-\lambda
q/m_{_{\rm P}}}\label{qeq} \eeq
is the adopted potential for the field $q$, $,q$ [dot] stands for
derivative w.r.t  $q$ [the cosmic time, $t$] and $H$ is the Hubble
expansion parameter,
\begin{equation}\label{rhoqi}
\mbox{\sf (a)}~~H= \sqrt{\rho_q +\rho_{_{\rm R}}+ \rho_{_{\rm
M}}}/\sqrt{3}m_{_{\rm P}}~~\mbox{with}~~\mbox{\sf
(b)}~~\rho_q=\frac{1}{2}\dot q^2+V,\eeq
the energy density of $q$ and $m_{_{\rm P}}=M_{\rm P}/8\pi$, where
$M_{\rm P}=1.22\times10^{19}~{\rm GeV}$ is the Planck mass. The
energy density of radiation, $\rho_{_{\rm R}}$, can be evaluated
as a function of temperature, $T$, whilst the energy density of
matter, $\rho_{_{\rm M}}$, with reference to its present-day
value:
\beq\label{rhos} \mbox{\sf (a)}~~\rho_{_{\rm
R}}=\frac{\pi^2}{30}g_{\rho*}\ T^4~~\mbox{and}~~\mbox{\sf
(b)}~~\rho_{_{\rm M}}R^3=\rho^0_{_{\rm M}}R_0^3
\end{equation}
with $R$, the scale factor of the universe. Assuming no entropy
production caused by the domination of another field (entropy
production due to the $q$ domination is not expected, since it
does not couple to matter), the entropy density, $s$, satisfies
the following two equations:
\beq \mbox{\sf (a)}~~sR^3=s_{\rm p}R_{\rm
p}^3~~\mbox{where}~~\mbox{\sf (b)}~~s=\frac{2\pi^2}{45}g_{s*}\
T^3, \label{rs}\eeq
where subscript ``p'' represents a specific reference point at
which the constant quantity $sR^3$ is evaluated and
$g_{\rho*}(T)~[g_{s*}(T)]$ is the energy [entropy] effective
number of degrees of freedom at temperature $T$. Their numerical
values are evaluated by using the tables included in {\tt
micrOMEGAs} \cite{micro}, originated from the {\sf DarkSUSY}
package \cite{dark} (recent improvements \cite{qcd} do not affect
essentially the results).

\subsubsection{Reformulation.} The numerical integration of eq.
(\ref{qeq}) is facilitated by converting the time derivatives to
derivatives w.r.t the logarithmic time \cite{brazil, german}:
\beq \vtau=\ln\left(R/R_0\right)=-\ln
(1+z)~~(\Rightarrow~\dot{\vtau}=H)\label{dtau} \eeq
with $z$ the redshift. Changing the differentiation, eq.
(\ref{qeq}) turns out to be equivalent to the system of two
first-order equations (prime denotes derivative w.r.t $\vtau$):
\beq \mbox{\sf (a)}~~Q=Hq^\prime~~\mbox{and}~~\mbox{\sf (b)}~~H
Q^\prime+3 H Q+ V_{,q}=0~~ \mbox{with}~~\mbox{\sf
(c)}~~\rho_q=\frac{1}{2}Q^2+V.\label{nsys}\eeq
In terms of $\vtau$ in eq. (\ref{dtau}), $s$ and $T$ can be
expressed through the relations :
\beq \mbox{\sf (a)}~~s=s_0 e^{-3\vtauf}~~\mbox{and}~~\mbox{\sf
(b)}~~ T=T_0 \left(\frac{g^0_{s*}}{g_{s*}}\right)^{1/3}
e^{-{\vtauf}}, \label{sTtau}\eeq
where eqs.~(\ref{rs}{\sf a}) and (\ref{rs}{\sf b}) have been used.
Similarly, $\rho_{_{\rm R}}$ and $\rho_{_{\rm M}}$ are elegantly
cast in the form:
\beq \label{rhotau}\mbox{\sf (a)}~~\rho_{_{\rm R}}=\rho^0_{_{\rm
R}}\frac{g_{\rho*}}{g^0_{\rho*}}
\left(\frac{g^0_{s*}}{g_{s*}}\right)^{4/3}e^{-4\vtauf}~~\mbox{and}~~\mbox{\sf
(b)}~~\rho_{_{\rm M}}=\rho^0_{_{\rm M}}e^{-3\vtauf}.\eeq
Eq.~(\ref{rhotau}{\sf a}) was extracted by inserting eq.
(\ref{sTtau}{\sf b}) in eq. (\ref{rhos}{\sf a}).
Eq.~(\ref{rhotau}{\sf b}) can be derived by combining
eq.~(\ref{dtau}) with eq.~(\ref{rhos}{\sf b}).

\subsubsection{Normalized form.} \label{nf} An
even more numerically ``robust'' \cite{german} form of eq.
(\ref{nsys}) can be achieved, if we introduce the following
dimensionless quantities:
\beq \label{vrhos}\mbox{\sf (a)}~~\vrho_{_{\rm M[R]}}=\rho_{_{\rm
M[R]}}/\rho^0_{\rm c },~~\mbox{\sf (b)}~~\vVo=V_o/\rho^0_{\rm
c}~~\mbox{and}~~\mbox{\sf (c)}~~\vq=q/\sqrt{3}m_{_{\rm P}}.\eeq
Employing these quantities, eq. (\ref{nsys}) can be re-written as:
\beq \mbox{\sf (a)}~~\vQ=\vH\vq^\prime ~~\mbox{and}~~\mbox{\sf
(b)}~~\vH\vQ^\prime+3\vH\vQ+\bar V_{,\bar
q}=0~~\mbox{with}~~\mbox{\sf (c)}~~\vH^2=\vrho_q+\vrho_{_{\rm
R}}+\vrho_{_{\rm M}}, \label{vH} \eeq
where the following quantities have been defined:
\beq \mbox{\sf (a)}~~\vV=\vVo e^{-\sqrt{3}\lambda
\vq},~~\vH=H/H_0,~~\vQ=Q/\sqrt{\rho^0_{\rm
c}}~~\mbox{and}~~\mbox{\sf
(b)}~~\vrho_q=\vQ^2/2+\vV.\label{vrhoq}\eeq
In our numerical calculation, we use the following values:
\beq \rho^0_{\rm c}=8.099\times10^{-47}h_0^2~{\rm GeV^4
}~~\mbox{and}~~H_0=2.13\times10^{-42}h_0~{\rm GeV},\eeq
with $h_0=0.72$. Also, $\vrho^0_{_{\rm M}}=0.29$ and
$T_0=2.35\times 10^{-13}~{\rm GeV}$. Substituting the latter in
eq.~(\ref{rhos}{\sf a}), we obtain $\vrho^0_{_{\rm
R}}=8.04\times10^{-5}$.

\subsubsection{Extracted quantities.} The solution of eqs.~(\ref{vH}{\sf a})
and (\ref{vH}{\sf b}) allows us to calculate some measurable
quantities which are used in order to test the quitessential model
against observations (see sec. \ref{reqq}). These are the density
parameters of the $q$-field, radiation and matter
\beq \label{omegas}\Omega_i=\rho_i/(\rho_q+\rho_{_{\rm
R}}+\rho_{_{\rm M}})=\vrho_i/\vH^2,~~\mbox{where}~~i=q,~{\rm
R~~\mbox{and}~~M},\eeq
respectively and the equation-of-state parameter (or barotropic
index) of the $q$-field, $w_q$,
\beq \label{wq} w_q=(\dot q^2/2-V)/(\dot
q^2/2+V)=1-\vV/2\vrho_q.\eeq

\subsubsection{Initial Conditions.} In
order to solve eq. (\ref{vH}) two initial conditions are to be
specified: These could be the values of $q$ and $q^\prime$ at an
initial $\vtau$, $\vti$ \cite{brazil}. We take $q(\vti)=0$
throughout our investigation, without any lose of generality. This
is because possible use of $\vq(\vti)=\vq_{_{\rm I}}\neq 0$ is
equivalent as if we had $\vq(\vti)=0$ and rescaled $\vVo$ to
$\vVo\exp{(-\sqrt{3}\lambda \vq_{_{\rm I}})}$ \cite{brazil}. This
displacement influences just the choice of $\vVo$ determined from
eq.~(\ref{rhoq0}).

On the other hand, the value of $q^\prime(\vti)$ is not a suitable
initial condition for our purposes. This is, because we wish to
focus on the regime (see eqs.~(\ref{omegas}), (\ref{vrhoq}{\sf b})
and (\ref{vH}{\sf a})):
\beq \Omega_q(\vti)\simeq\frac{\vH_{\rm I
}^2q^{\prime2}(\vti)/2}{\vH_{\rm I}^2}\simeq1~\Rightarrow~
q^\prime(\vti)\simeq\sqrt{2},\eeq
where we take $\vrho_{q_{\rm I}}=\vrho_q(\vti)\simeq\vQ^2(\vti)/2$
and $\vHi=\vH(\vti)$. This means that $\vQ_{\rm I}=\vQ(\vti)$
tends to infinity, since inserting eqs.~(\ref{vrhoq}{\sf b}) and
(\ref{vH}{\sf a}) into eq.~(\ref{vH}{\sf c}) we can obtain:
\beq \vQ=|\vq^\prime|\sqrt{\frac{\vV+\vrho_{_{\rm R}}+\vrho_{_{\rm
M}}}{1-\vq^{\prime\;2}/2}}\cdot\eeq
In order to handle properly this subtlety, we find it convenient
to define as initial condition, the square root of the
kinetic-energy density of $q$ at $\vti$,
\beq\sqrt{\vrho_{_{\rm
KI}}}=\vQ(\vti)/\sqrt{2}\simeq\sqrt{\vrho_{q_{\rm
I}}}\simeq\vHi.\eeq

\subsection{Q{\ssz UINTESSENTIAL} D{\ssz YNAMICS}} \label{dynq}

\hspace{.67cm} We can obtain a comprehensive and rather accurate
approach of the $q$ dynamics, following the arguments of
ref.~\cite{dimopoulos}. Namely, $q$ undergoes the following three
phases:

\subsubsection{Kination Dominated Phase.} During this phase, the evolution
of both the universe and $q$ is dominated by the kinetic-energy
density of $q$. Consequently, eq.~(\ref{vH}{\sf a}) reads:
\beq \label{defk} {\sf (a)
}~~\vH\vQ^\prime+3\vH\vQ=0~~\mbox{with}~~{\sf (b)
}~~\vH=\sqrt{\vrho_q}\simeq\vQ/\sqrt{2}.\eeq
The former equation can be integrated trivially, with result:
\beq \label{rhok} \vQ=\vQ_{\rm I}e^{-3(\vtauf-\vtif)}~\Rightarrow
~\rho_q=\rho_{q_{\rm I}}e^{-6(\vtauf-\vtif)}. \eeq
Combining eq. (\ref{defk}{\sf b}) with (\ref{vH}), we obtain:
\beq\label{qk} \vq\simeq\sqrt{2}\ (\vtau-\vti)~~(\Rightarrow
~\vq^\prime=\sqrt{2})~~\mbox{for}~~\vtau\leq\vtkr\eeq
with $\vtkr$, the point where the totally KD phase is terminated.
This occurs, when:
\beq \label{tkr} \rho_{_{\rm R}}(\vtau_{_{\rm
KR}})=\rho_q(\vtau_{_{\rm KR}})~\Rightarrow~\vtau_{_{\rm
KR}}\simeq\vtau_{_{\rm I}}+\ln\sqrt{\frac{\rho_{q_{\rm
I}}}{\rho_{_{\rm RI}}}},\eeq
where the right hand side of eq. (\ref{rhok}) has been equated to
the expression below at $\vtau=\vtkr$ (since $\vtau$ is close to
$\vti$ we suppose that $g_{\rho*}~[g_{s*}]$ does not vary from its
value at $\vti$, $g^{\rm I}_{\rho*}~[g^{\rm I}_{s*}]$):
\beq  \rho_{_{\rm R}}=\rho_{_{\rm RI}}\frac{g_{\rho*}}{g^{\rm
I}_{\rho*}} \left(\frac{g^{\rm
I}_{s*}}{g_{s*}}\right)^{4/3}e^{-4(\vtauf-\vtif)},
~~\mbox{with}~~\rho_{_{\rm RI}}=\rho_{_{\rm
R}}(\vti).\label{rhoi}\eeq
Eq.~(\ref{rs}{\sf a}) with reference point $\vti$ and
eq.~(\ref{rhos}{\sf a}) were employed in order to extract eq.
(\ref{rhoi}).

\subsubsection{Frozen-Field Dominated (FD) Phase.} For $\vtau>\vtkr$, the
universe becomes RD but the evolution of $q$ continues to be
dominated by its kinetic energy density. Therefore,
\beq \label{deff} {\sf (a)
}~~\vH\vQ^\prime+3\vH\vQ=0~~\mbox{with}~~{\sf (b)
}~~\vH=\sqrt{\vrho_{_{\rm R }}}.\eeq
Inserting eq.~(\ref{deff}{\sf b}) into eq. (\ref{vH}{\sf a}) and
integrating the resulting one, we obtain:
\beq\label{qf} \vq=\vq_{_{\rm KR}}+\vrho^{-1/2}_{_{\rm
RI}}\vQ_{\rm I}e^{-(\vtauf_{_{\rm KR}}
-\vtif)}\left(1-e^{-(\vtauf-\vtauf_{_{\rm
KR}})}\right)~~\mbox{for}~~\vtkr<\vtau\leq\vtau_{_{\rm FA}}\eeq
where $\vq_{_{\rm KR}}=\vq(\vtau_{_{\rm
KR}})=\ln\left(\rho_{q_{\rm I}}/\rho_{_{\rm RI}}\right)/\sqrt{2}$
from eqs.~(\ref{qk}) and (\ref{tkr}) and $\vtau_{_{\rm FA}}$ is
specified in eq.~(\ref{tfA}). It is obvious from eq.~(\ref{qf})
that $q$ freezes at about $\vtau_{_{\rm KF}}\simeq\vtau_{_{\rm
KR}}+6$ to the value:
\beq\label{qff} \vq_{_{\rm F}}\simeq\vq_{_{\rm
KR}}+\sqrt{2}~~(\Rightarrow
~\vq^\prime=0)~~\mbox{for}~~\vtau_{_{\rm
KF}}\leq\vtau\leq\vtau_{_{\rm FA}}. \eeq
Note that $\vrho_q$ reaches its constant value, $\vrho_{q_{\rm
F}}=\vV(q_{_{\rm F}})$, at $\vtau_{_{\rm PL}}\gg\vtau_{_{\rm KF}}$
such, that:
\beq \vQ^2(\vtau_{_{\rm PL}})/2=\vV(q_{_{\rm
F}})~\Rightarrow~\vtau_{_{\rm PL}}=\vti+\lambda\vq_{_{\rm
F}}/2\sqrt{3}-\ln(\vVo/\vrho_{q_{\rm I }})/6.\eeq
\subsubsection{Attractor Dominated (AD) Phase.} For $\vtau>\vtau_{_{\rm PL}}$,
$\rho_q$ becomes $V$ dominated as in the case of inflation.
Consequently, the evolution of $q$ is described by the following:
\beq \label{sysA} \mbox{\sf (a)}~~3\vH\vQ+\bar V_{,\bar
q}\simeq0~~\mbox{with}~~\mbox{\sf (b)}~~\vH=\sqrt{\vrho_{_{\rm
B}}}\simeq\sqrt{\vrho^0_{_{\rm B}}}e^{-3(1+w_{_{\rm
B}})\vtauf/2},\eeq
where $\vrho_{_{\rm B}}$ is the dominant background-energy density
of the universe with $w_{_{\rm B}}=1/3~[0]$ for the
RD~[matter-dominated (MD)] era. As can be shown \cite{wet}, and
has been extensively discussed \cite{brazil,german, cline,
dimopoulos}, the system in eq.~(\ref{qeq}) admits:
\begin{itemize}
\item[{\sf (i)}] A global attractor for $\lambda>\sqrt{3(1+w_{_{\rm B}})}$ with
a fixed-point equation-of-state parameter $w^{\rm fp}_q=w_{_{\rm
B}}$ and density parameter $\Omega^{\rm fp}_q=3(1+w_{_{\rm
B}})/\lambda^2$. This is the so-called self-tuning
\cite{attr,expo, ferreira} case where the $q$ evolution is
insensitive to the choice of $\vVo$. However, this case can be
discarded \cite{brazil, german} since, it fails to meet the
observational data (see sec.~\ref{reqq}).

\item[{\sf (ii)}] A late time attractor for $\lambda<\sqrt{3(1+w_{_{\rm B}})}$ with
$w^{\rm fp}_q=\lambda^2/3-1$ and $\Omega^{\rm fp}_q=1$. As is
pointed out \cite{brazil, german} and we verify in
sec.~\ref{Qpar}, the model can satisfy a number of observational
constraints, for a reasonable set of initial conditions.
\end{itemize}
Inserting eq.~(\ref{vH}{\sf a}) into eq.~(\ref{sysA}{\sf a}) and
integrating the resulting equation, we obtain for the latter case:
\beq\label{qA}
\vq=\lambda\vtau/\sqrt{3}+\ln(\vVo/\vrho^0_q)/\sqrt{3}\lambda
~~(\Rightarrow
~\vq^\prime=\lambda/\sqrt{3})~~\mbox{for}~~\vtau>\vtau_{_{\rm
FA}}\eeq
where the transition from the FD to the AD phase occurs at the
point $\vtau_{_{\rm FA}}$, which can be estimated by:
\beq\label{tfA}\vtau_{_{\rm
FA}}=\sqrt{6}/\lambda+\sqrt{3}\vq_{_{\rm
KR}}/\lambda-\ln(\vVo/\vrho^0_q)/\lambda^2. \eeq
The latter can be easily extracted by equating the values of the
expressions in eqs.~(\ref{qf}) and (\ref{qA}) for
$\vtau=\vtau_{_{\rm FA}}$.
Employing eq.~(\ref{qA}), we can derive $Q$ via eq.~(\ref{vH}{\sf
a}). Inserting it in the relation $\vrho_q=\vQ^2/(1+w_q)$, which
can be derived from eq.~(\ref{wq}), we arrive at the energy
density of the late-time attractor:
\beq\vrho_{_{\rm A}}\simeq\vrho_q^0 e^{-3(1+w^{\rm
fp}_q)\vtauf}~~\mbox{with}~~w^{\rm fp}_q=\lambda^2/3-1.
\label{attr}\eeq

\subsection{Q{\ssz UINTESSENTIAL} R{\ssz EQUIREMENTS}} \label{reqq}

\hspace{.67cm}  We briefly describe the various criteria that we
impose on our quintessential model.
\subsubsection{KD ``Constraint''.} For the purposes of the present paper,
we desire to focus our attention on the range of parameters which
ensure an absolute  [at least relative domination]  of the
$q$-kinetic energy at $\vti$. This can be achieved, when:
\beq\mbox{\sf (a)}~~\Omega_q(\vti)=1~~\left[\mbox{\sf
(b)}~~0.5\leq\Omega_q(\vti)~~\mbox{and}~~\mbox{\sf (c)}
~~\Omega_q(\vti)<1\right].\label{domk}\eeq
Ranges of parameters, which meet all the residual constraints of
this section, not restricted by eq.~(\ref{domk}) are explored in
ref.~\cite{brazil}.
\subsubsection{Nucleosynthesis (NS) Constraint.} The
presence of $\vrho_q$ has not to spoil the successful predictions
of Big Bang NS which commences at about $\vtns=-22.5$
corresponding (see eq.~(\ref{sTtau}{\sf b})) to $T_{\rm NS}=1~{\rm
MeV}$ \cite{oliven}. Taking into account the most up-to-date
analysis of ref.~\cite{oliven}, we adopt a rather conservative
upper bound on $\Omega_q(\vtns)$, less restrictive than that of
ref.~\cite{nsb}. Namely, we require:
\beq\Omega_q^{\rm NS}=\Omega_q(\vtns)\leq0.21~~\mbox{($95\%$
c.l.)} \label{nuc}\eeq
which corresponds to additional effective neutrinos species
$\delta N_\nu<1.6$ \cite{oliven}. Let us note that extra
contribution in the left hand side of eq.~(\ref{nuc})  due to
energy density of gravitational waves (GWs) created during the
transition from the KD to RD era is negligible as we infer by
explicitly applying the formulae of ref.~\cite{miai}. On the other
hand, we do not consider contributions (potentially large
\cite{giova}) due to GWs generated during a possible former
transition from inflation to KD epoch. The reason is that
inflation could be driven by another field different to $q$ and
so, any additional constraint arisen from this period would be
highly model dependent.

\subsubsection{Coincidence Constraint.} The present value of
$\vrho_q$, $\vrho^0_q$, must be compatible with the preferred
range of eq.~(\ref{cdmba}{\sf b}). This can be achieved by
adjusting the value of $\vVo$. Since, this value does not affect
crucially our results (especially on the CDM abundance), we decide
to fix $\vrho^0_q$ to its central experimental value, demanding:
\beq \Omega^0_q=\vrho^0_q=0.73.\label{rhoq0}\eeq
\subsubsection{Acceleration Constraint.} A successful
quintessential scenario has to account for the present-day
acceleration of the universe, i.e. \cite{wmap},
\beq -1\leq w_q(0)\leq-0.78~~\mbox{($95\%$ c.l.)}.\label{wq0}\eeq
In addition, since the string theory disfavors the eternal
acceleration, it would be desirable to demand $w^{\rm fp}_q>-1/3$
\cite{dimopoulos}. However, in the case of the used potential, we
did not succeed to achieve compatibility of the latter optional
restriction with eq.~(\ref{wq0}), in accordance with the findings
of ref.~\cite{german}.

\subsubsection{Residual Constraints.} In our scanning, finally, we take
into account the following less restrictive but also non-rigorous
bounds, which, however, do not affect crucially our results:
\beq {\sf (a)}~~-50\lesssim  \vti\lesssim-36~~\mbox{and}~~{\sf
(b)}~~\bar H_{\rm I}\lesssim 10^{56}. \label{para}\eeq
The lower bound of eq.~(\ref{para}{\sf a}) comes from the
gravitino constraint \cite{gravitinoc} which provides an upper
bound on the reheat temperature, $T_{\rm RH}<(10^{9}-10^{10})~{\rm
GeV}$. This can be translated to a lower bound on $\vti$, through
eq.~(\ref{sTtau}{\sf b}). However, this bound may not be so
reliable, since there is no thorough investigation of the
gravitino constraint within the context of quintessential
cosmology, to date. Also, since we do not study the evolution of
the universe before the commencement of the KD era, we wish to
liberate our $\Omega_{\chi}h_0^2$ calculation from this ignorance.
To this end, we demand $\vti$ to be lower than the upper bound of
eq.~(\ref{para}{\sf a}). This corresponds to the onset of the
Boltzmann suppression of the $\chi$-number density (see
sec.~\ref{Neqs}) for mass of $\chi$ equal to $500~{\rm GeV}$ (see
eq.~(\ref{mchi})). The bound of eq.~(\ref{para}{\sf b}) comes from
the COBE constraints \cite{cobe} on the spectrum of GWs produced
at the end of inflation \cite{qinfHI}.

\section{CDM A{\ftn BUNDANCE IN THE} P{\ftn RESENCE OF THE} KD
P{\ftn HASE}} \label{cdmhr}\setcounter{equation}{0}

\hspace{.67cm} We assume that the CDM candidate, $\chi$, maintains
kinetic and chemical equilibrium (see below) with plasma, is
produced through thermal scatterings and decouples (being
non-relativistic) during the KD epoch. Our theoretical analysis is
presented in sec. \ref{Beq} and its numerical treatment in sec.
\ref{Neqs}. Useful approximated expressions are derived in sec.
\ref{Seqs}.

\subsection{T{\ssz HE} B{\ssz OLTZMANN} E{\ssz QUATION}}
\label{Beq}

\hspace{.67cm} Since the $\chi$ particles are in kinetic
equilibrium with the cosmic fluid, their number density,
$n_{\chi}$, satisfies the following Boltzmann equation:
\beq \dot n_\chi+3Hn_\chi+\langle \sigma v \rangle \left(n_\chi^2
- n_\chi^{\rm eq2}\right)=0,\label{nx}\eeq
where $H$ is given by eq. (\ref{rhoqi}{\sf a}), $\langle \sigma v
\rangle$ is the thermal-averaged cross section of $\chi$ particles
times the velocity and $n_{\chi}^{\rm eq}$ is the equilibrium
number density of $\chi$, which obeys the Maxwell-Boltzmann
statistics:
\begin{equation} \label{neq}
n_{\chi}^{\rm eq}(x)=\frac{g}{(2\pi)^{3/2}}
m_{\chi}^3\>x^{3/2}\>e^{-1/x}P_2(1/x),~~\mbox{where}~~
x=T/m_{\chi}
\end{equation}
with $m_\chi$ the mass of $\chi$. We pose $g=2$ for the number of
degrees of freedom of ${\chi}$ and $P_n(z)=1+(4n^2-1)/8z$ is
obtained by asymptotically expanding the modified Bessel function
of the second kind of order $n$. Note that
non-chemical-equilibrium production of $\chi$'s requires
$\sv<10^{-20}~{\rm GeV}^{-2}$ \cite{fornengo}. Since such a value
is well below the usually obtainable values \cite{lkk, lzk,
branon, fornengo}, we do not consider further this possibility,
here.

\subsection{N{\ssz UMERICAL} C{\ssz ALCULATION}} \label{Neqs}

\hspace{.67cm} Following the strategy of sec. \ref{nf}, we
introduce the dimensionless quantities:
\beq \vrho^{[\rm eq]}_\chi=\rho^{[\rm eq]}_\chi/\rho^0_{\rm
c},~~\mbox{where}~~\rho^{[\rm eq]}_\chi=m_\chi n^{[\rm
eq]}_\chi.\eeq
In terms of these, eq. (\ref{nx}) takes the following master, for
numerical manipulations, form :
\beq \vH\vrho^\prime_\chi+3\vH \vrho_{\chi}+\vsv
\left(\vrho_{\chi}^2 - \vrho_{\chi}^{\rm
eq2}\right)/\vm_\chi=0,\label{rx}\eeq
where $\vH$ is given by eq.~(\ref{vH}{\sf c}) and the following
quantities have been defined:
\beq \vsv=\sqrt{\rho^0_{\rm c}} \langle \sigma v
\rangle~~\mbox{and}~~\vm_\chi=m_\chi/\sqrt{3}m_{_{\rm P}}.\eeq
Eq.~(\ref{rx}) can be solved numerically  with initial condition
$\vrho_\chi(\vtb)=\vrho^{\rm eq }_\chi(\vtb)$, where $\vtb$
corresponds (see eq.~(\ref{sTtau}{\sf b})) to the beginning
($x=1$) of the Boltzmann suppression of $\vrho_{\chi}^{\rm eq}$.
Since $\vti<\vtb$, the integration of eq.~(\ref{rx}) is realized
from $\vtb$ down to $0$. Finally, $\Omega_{\chi}h_0^2$ can be
easily found, via the relation:
\beq \label{omega} \Omega_{\chi}h_0^2=\vrho_{\chi}(0)h_0^2.\eeq

\subsection{S{\ssz EMI}-A{\ssz NALYTICAL}
C{\ssz ALCULATION}} \label{Seqs}

\hspace{.67cm} The aim of this section is the calculation of
$\Omega_{\chi}h_0^2$ based on the already obtained semi-analytical
expressions of sec. \ref{dynq}. The procedure is described
step-by-step below.

\subsubsection{Reformulation of the Boltzmann Equation.}
Introducing the variables $Y^{\rm [eq]}=n^{\rm [eq]}_{\chi}/s$
\cite{kolb, gelmini}  (in order to absorb the dilution term) and
converting the derivatives w.r.t $t$, to derivatives w.r.t
$\vtau$, eq.~(\ref{nx}) can be rewritten as:
\beq \label{BE} HY^\prime=-\sv\left(Y^2 -Y^{\rm eq2} \right)s,\eeq
where eq. (\ref{sTtau}{\sf a}) has been also utilized.
Substituting eqs.~(\ref{rhoi}) and (\ref{rhok}) in eq.
(\ref{rhoqi}{\sf a}) and ignoring the negligible, during the
$\chi$ decoupling, contribution of $\rho_{_{\rm M}}$, $H$ can be
expressed as:
\bea \label{gq} \mbox{\sf (a)}~~H=\sqrt{\rho_{_{\rm
R}}g_q/3m^2_{\rm P }},~~\mbox{where}~~\mbox{\sf
(b)}~~g_q=1+r_q~~\mbox{with}\\ \label{rq} r_q=r_{_{\rm
I}}\frac{g^{\rm I}_{\rho*}}{g_{\rho*}}\left(\frac{g_{s*}}{g^{\rm
I}_{s*}}\right)^{4/3}e^{-2(\vtauf-\vtif)}~~\mbox{and}~~r_{_{\rm
I}}=\frac{\rho_{q_{\rm I}}}{\rho_{_{\rm RI}}}. \eea
Equivalently for $\vtau_{_{\rm PL}}>\vtns$, taking as reference
point $\vtns$ instead $\vti$ in eqs.~(\ref{rhoi}) and
(\ref{rhok}), we obtain:
\beq \label{rqns} r_q=r_{_{\rm NS}}\frac{g^{\rm
NS}_{\rho*}}{g_{\rho*}}\left(\frac{g_{s*}}{g^{\rm
NS}_{s*}}\right)^{4/3}e^{-2(\vtauf-\vtauf_{_{\rm NS}}
)}~~\mbox{with}~~r_{_{\rm \rm NS}}=\frac{\rho_{q}^{\rm
NS}}{\rho_{_{\rm R}}^{\rm NS}}=\frac{\Omega^{\rm
NS}_q}{1-\Omega^{\rm NS}_q}\eeq
and the superscript NS denotes the values of the several
quantities at $\vtns$. Inserting eqs.~(\ref{gq}),
(\ref{rhotau}{\sf a}) and (\ref{sTtau}{\sf a}) into
eq.~(\ref{BE}), this can be cast in the following final form:
\bea Y^{\prime}= \frac{y\ \sv}{\sqrt{g_q}} \left(Y^2 -Y^{\rm
eq2}\right),~~\mbox{where}\label{BEf}\\
y(\vtau)=-\frac{s_0}{H^{0}_{\rm R}}\left(\frac{g^{\rm
0}_{\rho*}}{g_{\rho*}}\right)^{1/2}\left(\frac{g_{s*}}{g^{\rm
0}_{s*}}\right)^{2/3} e^{-\vtauf}~~\mbox{with}~~H^0_{\rm
R}=\sqrt{\rho^{0}_{_{\rm R}}/3m^2_{\rm P}}.\label{ytau}\eea
\subsubsection{The freeze-out procedure.\label{xfsec}} In the case
of the equilibrium $\chi$ production, an accurately approximate
solution of eq.~(\ref{BEf}) can be achieved, introducing the
notion of freeze-out temperature, $T_{\rm F}=T(\vtf)=x_{_{\rm
F}}m_{\chi}$ \cite{kolb, gelmini}, which allows us to study
eq.~(\ref{BEf}) in the two extreme regimes:

$\bullet$ At very early times, when $\vtau\ll \vtf$, $\chi$'s are
very close to equilibrium. So, it is more convenient to rewrite
eq.~(\ref{BEf}) in terms of the variable
$\Delta(\vtau)=Y(\vtau)-Y^{\rm eq}(\vtau)$ as follows:
\beq \label{deltaBE} \Delta^{\prime}=-{Y^{\rm eq}}^{\prime}+y\ \sv
\Delta\left(\Delta+2Y^{\rm eq}\right)/\sqrt{g_q}. \eeq
The freeze-out point $\vtf$ can be defined by
\beq \Delta(\vtf)=\delta_{\rm F}Y^{\rm eq}(\vtf)~~\Rightarrow
~~\Delta(\vtf)\Big(\Delta(\vtf)+2Y^{\rm eq}(\vtf)\Big)=\delta_{\rm
F}(\delta_{\rm F}+2)\ Y^{\rm eq2}(\vtf), \label{Tf} \eeq
where $\delta_{\rm F}$ is a constant of order one, determined by
comparing the exact numerical solution of eq.~(\ref{BEf}) with the
approximate under consideration one. Inserting eqs.~(\ref{Tf})
into eq.~(\ref{deltaBE}), we obtain the following equation, which
can be solved w.r.t $\vtf$ iteratively:
\bea \Big(\ln Y^{\rm eq}\Big)^\prime(\vtf) =y_{_{\rm
F}}\sv\delta_{\rm F} (\delta_{\rm F}+2) Y^{\rm
eq}(\vtf)/\sqrt{g_q}(\delta_{\rm F}+1)~~\mbox{with} \label{xf}
\\ \label{xfa}y_{_{\rm F}}= y(\vtf)
~~\mbox{and}~~\Big(\ln Y^{\rm
eq}\Big)^{\prime}(\vtau)=x^\prime\left(\frac{1}{x^2}-\frac{3}{2x}-
\frac{g^{\prime}_{s*}}{g_{s*}}+\frac{15}{8P_2(1/x)}\right)\cdot
\eea

$\bullet$ At late times, when $\vtau\gg\vtf$, $Y\gg Y^{\rm eq}$
and so, $Y^2-Y^{\rm eq2}\simeq Y^2$. Inserting this into
eq.~(\ref{BEf}) and integrating the resulting equation from $\vtf$
down to 0, we arrive at:
\bea \label{BEsol} \mbox{\sf (a)}~~Y_0 = \left(Y_{\rm
F}^{-1}+J_{\rm F} \right)^{-1},~~\mbox{where}~~\mbox{\sf
(b)}~~J_{\rm F}= \int_{0}^{\vtauf_{_{\rm F}}} d\vtau\ \frac{y\
\sv}{\sqrt{g_q}}~~\mbox{and}~~\\
\label{BEyf} Y_{\rm F} =(\delta_{\rm F} +1)\> Y^{\rm
eq}(\vtf)~~\mbox{with}~~ Y^{\rm
eq}(\vtau)=\frac{g}{g_{s*}}\frac{45}{4\pi^4}\sqrt{\frac{\pi}{2}}
x^{-3/2}\ e^{-1/x}\ P_2(1/x), \eea
where the $x-\vtau$ dependence can be derived from
eq.~(\ref{sTtau}{\sf b}). Although not crucial, a  choice
$\delta_{\rm F}=1.2\mp0.2$ assists us to approach better the
precise numerical solution of eq.~(\ref{BEf}).

\subsubsection{The CDM abundance.} Our final aim is
the calculation of the current $\chi$ relic density, which is
based on the well known formula \cite{gelmini}:
\begin{equation}
\label{om1} \Omega_{\chi}=\rho_{\chi}^0/\rho_{\rm c}^0= m_{\chi}
s_0 Y_0/\rho_{\rm c}^0~\Rightarrow~\Omega_{\chi}h_0^2 = 2.741
\times 10^8\ Y_0\ m_{\chi}/\mbox{GeV}.
\end{equation}
The presence of $g_q>1$ in eq. (\ref{xf}) and, mainly, in
eq.~(\ref{BEsol}{\sf b}) reduces $J_{\rm F}$, thereby increasing
the value $\Omega_{\chi}h_0^2$ w.r.t the one obtained in the SC
(i.e. with $g_q=1$), $\Omega_{\chi}h_0^2|_{_{\rm SC}}$. The
resulting enhancement can be estimated, by defining the quantity
\cite{prof}:
\beq\label{dom}\Delta\Omega_{\chi}=
\Big(\Omega_{\chi}h_0^2-\Omega_{\chi}h_0^2|_{_{\rm
SC}}\Big)/\Omega_{\chi}h_0^2|_{_{\rm SC}}.\eeq

\subsubsection{The variation of $\Delta\Omega_{\chi}$.} \label{detvar} The
variation of $\Delta\Omega_{\chi}$ w.r.t the free parameters can
be designed by simplifying the formulas above. In particular,
$\vtf$ and $\Delta\Omega_{\chi}$ can be roughly estimated as:
\beq \label{delt}\kern-40.pt \mbox{\sf
(a)}~~\vtf\sim-2\ln(\sqrt{m_\chi}/\sv) ~~\mbox{and}~~\mbox{\sf
(b)}~~ \Delta\Omega_\chi\sim J_{\rm F}|_{_{\rm SC }}/J_{\rm
F}-1\sim \sqrt{g_q}-1\sim e^{-\vtauf_{_{\rm F}}} \hfill \eeq
where we kept only the most important terms in eqs.~(\ref{xf}),
(\ref{xfa}) and (\ref{ytau}). Also, we have taken into account
eqs.~(\ref{BEsol}{\sf b}), from which we extracted $J_{\rm
F}\sim\sv e^{-\vtauf_{_{\rm F}}}/\sqrt{g_q},~J_{\rm F}|_{_{\rm
SC}}\sim\sv e^{-\vtauf_{_{\rm F}}}$ (for constant $\sv=a$).

Armed with these formulas, we can explain that
$\Delta\Omega_{\chi}$ increases as: {\sf (i)} $g_q$ increases~(for
fixed $m_\chi$ and $\sv$); this is obvious from
eq.~(\ref{delt}{\sf b}). {\sf (ii)} $\sv$ decreases (for fixed
$m_\chi$ and $g_q$); indeed, from eq.~(\ref{delt}{\sf a}),
decrease of $\sv$ results to a decrease of $\vtf$, which in turn,
causes an increase of $\Delta\Omega_\chi$. {\sf (iii)} $m_\chi$
increases (for fixed $\sv$ and $g_q$); indeed, as shown from
eq.~(\ref{delt}{\sf a}), an increase of $m_\chi$ generates a
decrease of $\vtf$ which increases $\Delta\Omega_\chi$.

\section{A{\ssz PPLICATIONS}} \label{ap}\setcounter{equation}{0}

\hspace{.67cm} Our numerical investigation depends on the
parameters:
$$\lambda,\ \vti,\ \vHi,\ m_{\chi},\ \sv .$$
For ease of reference we call the three first parameters $q$
parameters, whereas the two later, CDM parameters. Recall that we
use $q(\vti)=0$ throughout and \vVo\ is adjusted so that
eq.~(\ref{rhoq0}) is satisfied. Nonetheless, for definiteness and
clarity, we give the used value of \vVo\ in the explicit examples
of figs. \ref{figw}, \ref{figr} and \ref{prepost}. In general,
$\vVo$ ranges between about 1 and $10^{33}$, increases with
$\lambda$ or $\vHi$ and turns out to be $(\vti,\vHi)$-independent,
for fixed $\lambda$ and $\Omega_q(\vti)<1$.

As regards the CDM parameters, we have to clarify that \sv\ can be
derived from $m_{\chi}$ and the residual (s)-particle spectrum,
once a specific theory has been adopted. However, to keep our
presentation as general as possible, we decide to treat $m_{\chi}$
and \sv\ as unrelated input parameters (following our strategy in
ref.~\cite{snr}). Specifically, keeping in mind that the most
promising CDM particle is the LSP, we focus our attention on the
range:
\beq 200~{\rm GeV} \leq m_{\chi}\leq500~{\rm GeV}.\label{mchi}\eeq
Taking into account the experimental constraints on the SUSY
spectra of several SUSY models (see, e.g., fig. 23 of
ref.~\cite{wmapl}), we adopt a rather restrictive lower bound on
$m_\chi$ which, however, ensures us that the range of
eq.~(\ref{mchi}) is valid even in the most constrained cases. The
upper bound in eq.~(\ref{mchi}) is imposed in order the analyzed
range to be possibly detectable in the future experiments (see,
e.g. ref.~\cite{munoz}). On the other hand, we isolate the two
extreme cases which we encounter when we use the non-relativistic
expansion in order to calculate $\sv$ (the method gives, in
general, accurate results far enough from $s$-poles and thresholds
\cite{gelmini, lah}):
\beq {\sf (a)}~~\sv=a~~\mbox{or}~~{\sf (b)}~~\sv=bx.
\label{sv}\eeq
The $x$ dependence in eq.~(\ref{sv}{\sf b}) emerges in the case of
a bino LSP \cite{Cmssm} without coannihilations (CANs), whereas
eq.~(\ref{sv}{\sf a}) is extracted in the majority of the residual
cases \cite{lkk}-\cite{branon}, \cite{ellis2}-\cite{lah}. The
values of $a$ and $b$ can be restricted by applying the bounds of
eq.~(\ref{cdmb}). Comments on the naturalness of the required
values are given in sec.~\ref{svmp}.

The presentation of our results begins with the description of the
evolution of the various quintessential quantities in
sec.~\ref{Qev}. In sec.~\ref{Qpar}, we present the ranges of the
quintessential parameters, allowed by the constraints of
sec.~\ref{reqq}. In sec.~\ref{omenh}, we investigate the behaviour
of the $\Omega_\chi h_0^2$ enhancement and finally, in
sec.~\ref{NTR} we present areas compatible with eqs.~(\ref{cdmb}).

\subsection{E{\ssz VOLUTION OF THE}
Q{\ssz UINTESSENTIAL} Q{\ssz UANTITIES}} \label{Qev}

\begin{figure}[t]
\hspace*{-.25in}
\begin{minipage}{8in}
\epsfig{file=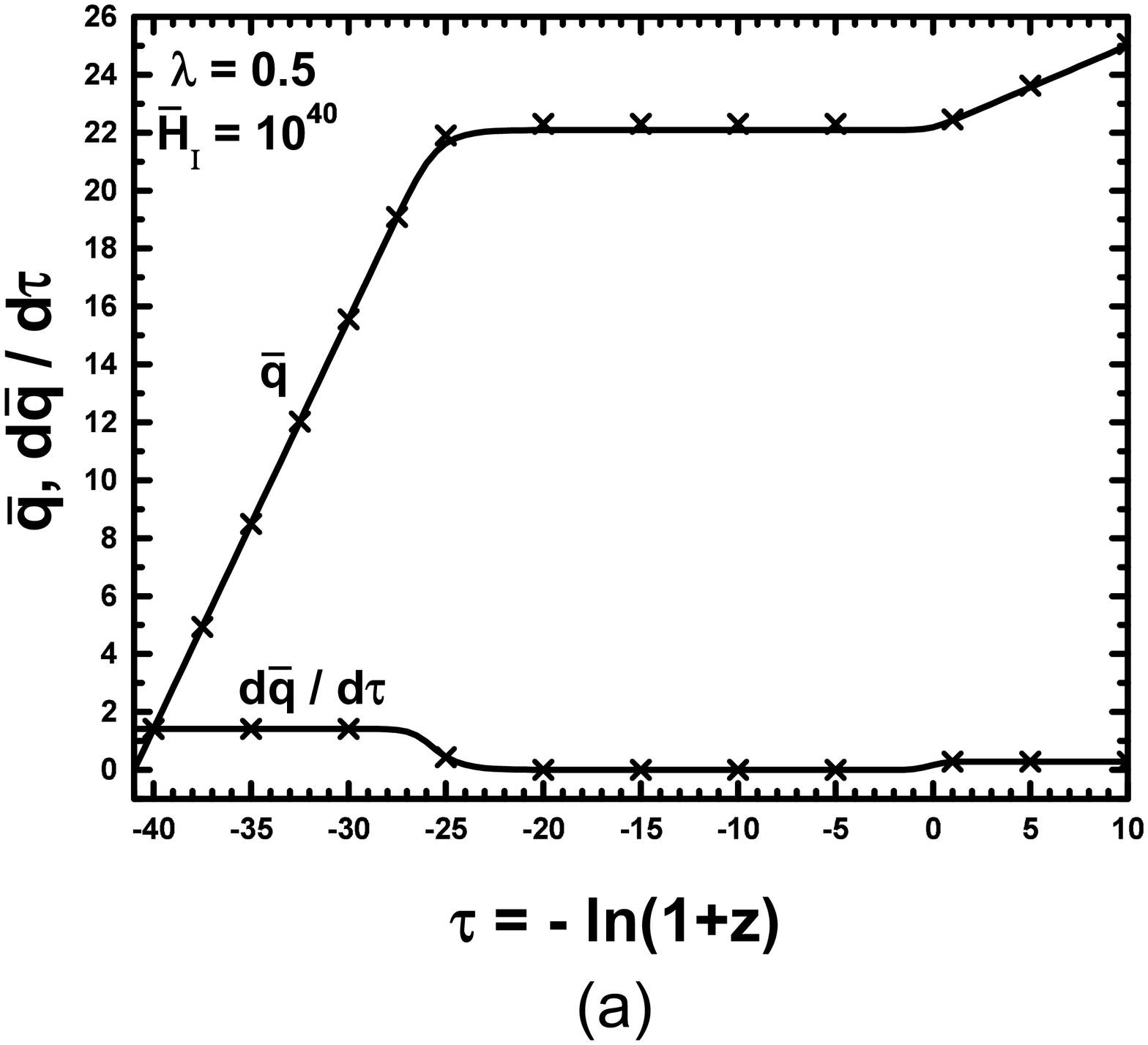,height=3.65in,angle=-90} \hspace*{-1.37 cm}
\epsfig{file=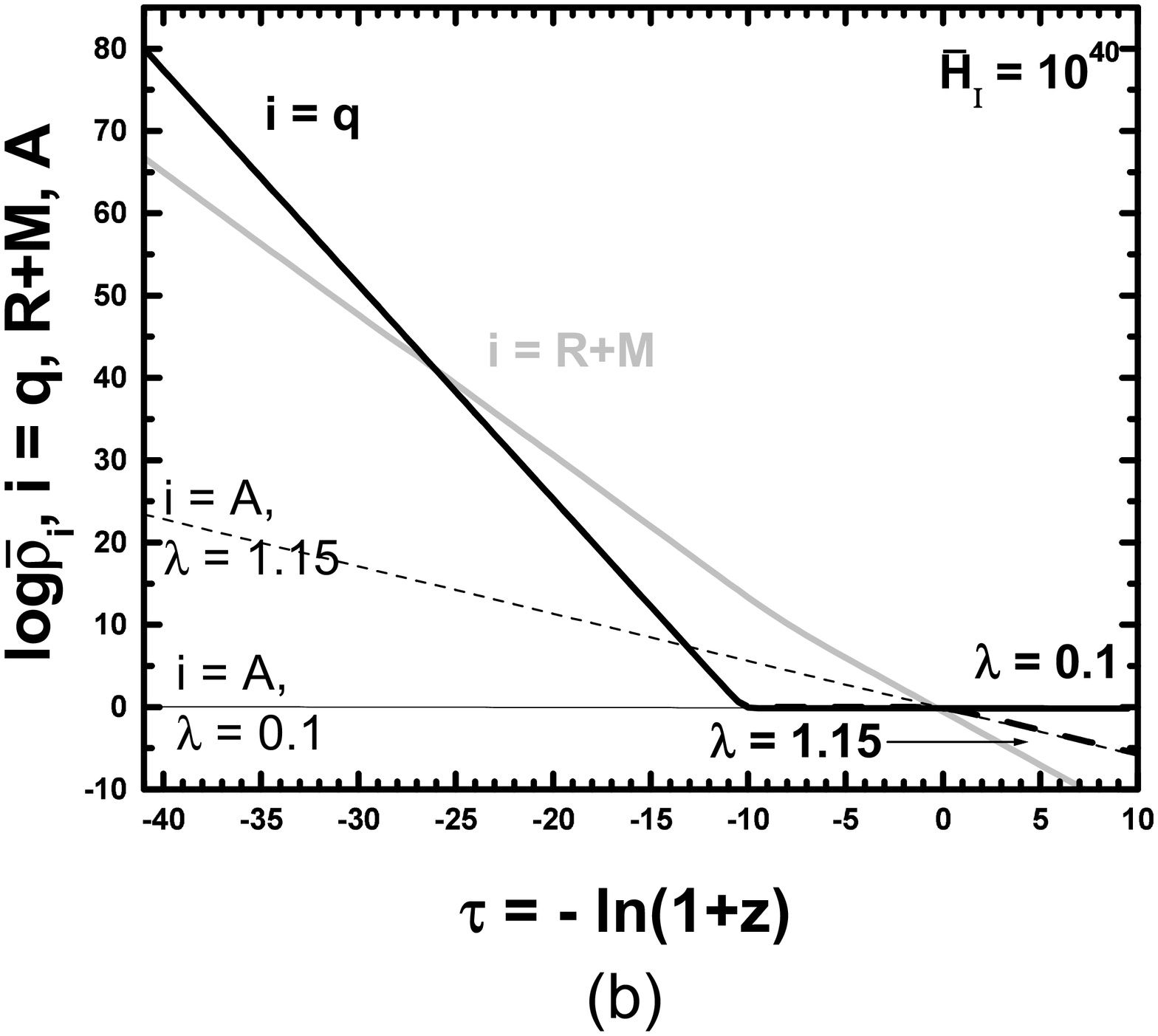,height=3.65in,angle=-90} \hfill
\end{minipage}\vspace*{-.01in}
\hfill\begin{center} \epsfig{file=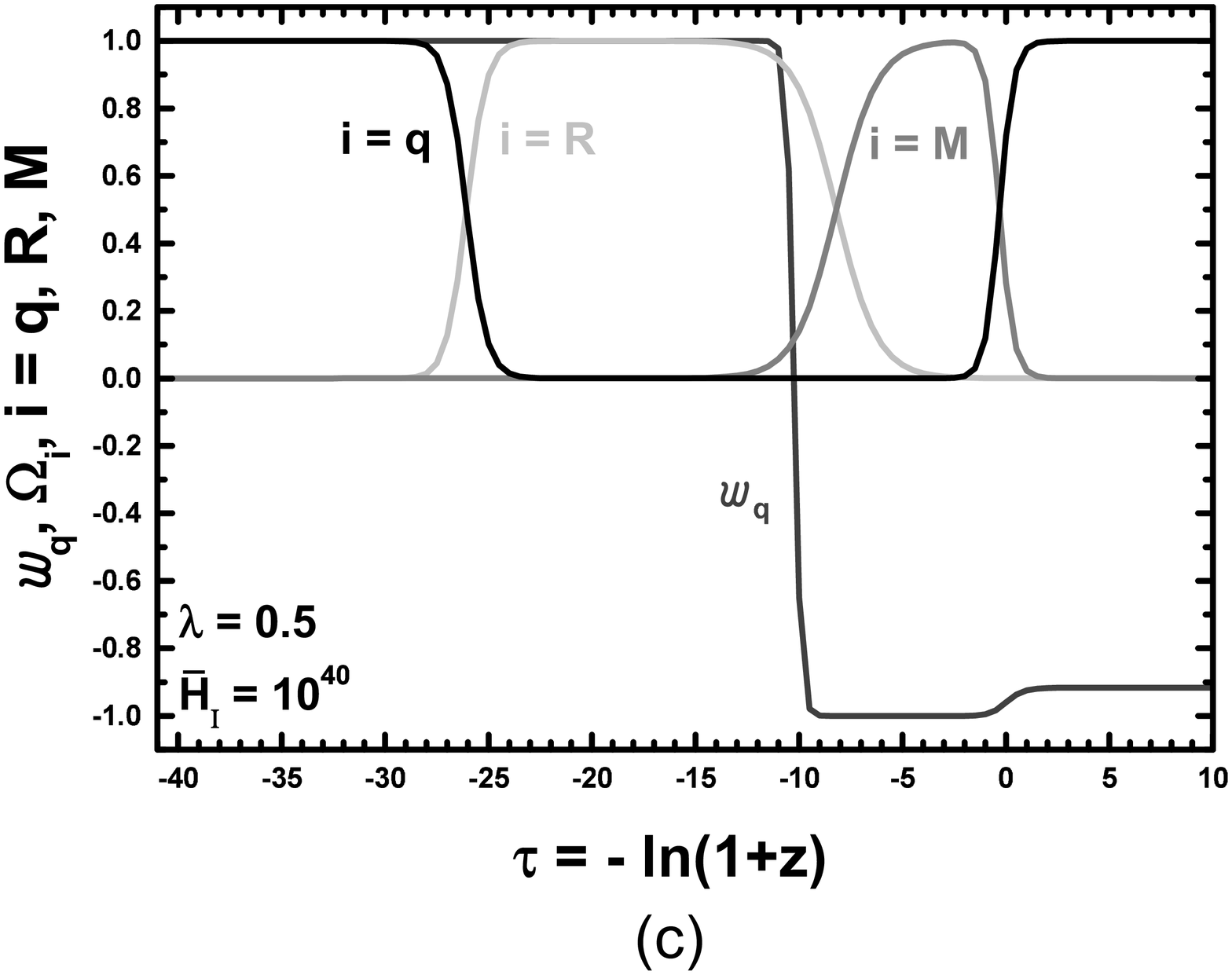,height=3.65in,angle=-90}
\end{center}
\hfill \caption[]{\sl The evolution as a function of $\vtauf$ for
$\vtif=-41$ and $\vHi=10^{40}$ of the quantities: $q$ and
$q^\prime$ for $\lambda=0.5 ~(\vVo=1.6\times 10^8)$  (crosses are
obtained by our analytical expressions) {\sf (a)}, $\log\vrho_i$
with $i=q$ (bold black lines), A (thin black lines) and R+M (light
grey line) for $\lambda=0.1~(\vVo=33.5)$ (solid lines) or
$\lambda=1.15~(\vVo=1.35\times10^{19})$ (dashed lines) {\sf (b)},
$w_q$ (dark gray line) and $\Omega_i$ with $i=q$ (black line), R
(light gray line) and M (gray line) for $\lambda=0.5
~(\vVo=1.6\times 10^8)$ {\sf (c)}.} \label{figw}
\end{figure}

\hspace{.67cm} We illustrate the evolution of the various
quintessential quantities presenting diagrams where in the
$x$-axis, $\vtau=-\ln(1+z)$ varies from $\vti$ down to late times,
e.g. 10 \cite{brazil, german, rosati}.

In fig. \ref{figw}-{\sf (a)}, we display $q$ and $q^\prime$ versus
$\vtau$ for $\vti=-41$, $\vHi=10^{40}$ and
$\lambda=0.5~(\vVo=1.6\times 10^8)$. Solid lines [crosses] are
obtained by numerically solving eq.~(\ref{vH}) [applying the
analytical expressions of sec.~\ref{dynq}]. Despite their
simplicity, our semi-analytical expressions in eqs.~(\ref{qk}),
(\ref{qf}) and (\ref{qA}), reproduce impressively the numerical
evolution of $q$ and $q^\prime$.

In fig. \ref{figw}-{\sf (b)}, we draw $\log\vrho_i$ versus $\vtau$
for $\vti=-41$, $\vHi=10^{40}$ and two ``extreme'' (see
figs.~\ref{ltH}-{\sf (a)} and {\sf (b)}) values of $\lambda$,
$\lambda=0.1~(\vVo=33.5)$ (solid lines) or
$\lambda=1.15~(\vVo=1.35\times10^{19})$ (dashed lines). For $i=q$
(bold black lines), we show $\log\vrho_q$, computed by inserting
in eq.~(\ref{vrhoq}{\sf b}) the numerical solution of
eqs.~(\ref{vH}{\sf a}) and (\ref{vH}{\sf b}). For $i={\rm A}$
(thin black lines), we show $\log\vrho_{_{\rm A}}$ derived from
eq.~(\ref{attr}). For $i={\rm R+M}$ (light grey line), we show
$\log\vrho_{_{\rm R+M}}$, which is the logarithm of the sum of the
contributions given by eqs.~(\ref{rhotau}{\sf a}) and
(\ref{rhotau}{\sf b}).

In fig.~\ref{figw}-{\sf (c)}, we plot $w_q$ (dark gray line) and
$\Omega_i$ with $i=q$ (black line), R (light gray line) and M
(gray line) versus $\vtau$, for $\vti=-41$, $\vHi=10^{40}$ and
$\lambda=0.5~(\vVo=1.6\times 10^8)$. The $y$-axis quantities are
computed by inserting in eqs.~(\ref{wq}) and (\ref{omegas}) the
numerical values obtained by eqs.~(\ref{vrhoq}{\sf a}),
(\ref{vrhoq}{\sf b}) and (\ref{rhotau}).

\begin{figure}[t]
\hspace*{-.25in}
\begin{minipage}{8in}
\epsfig{file=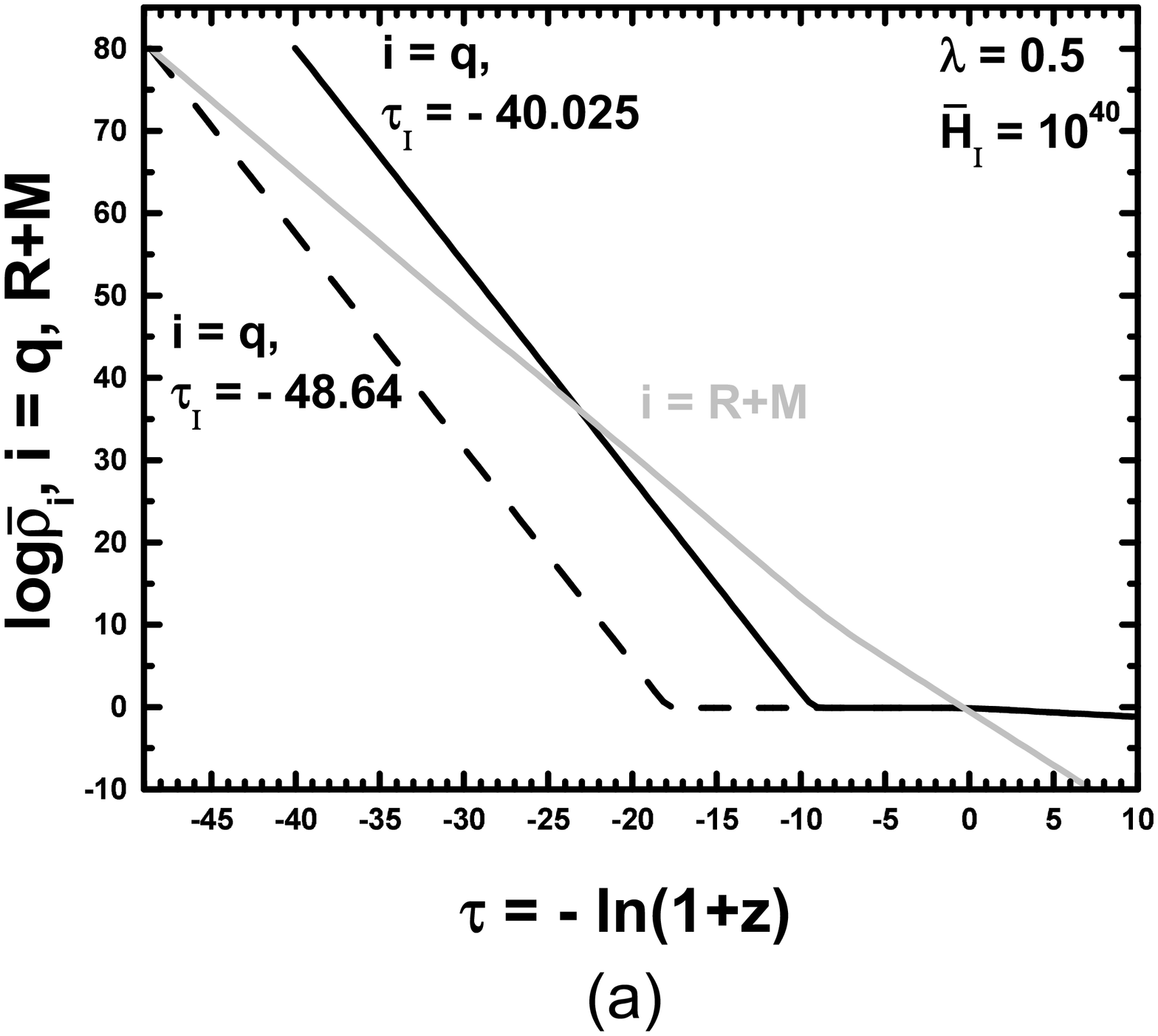,height=3.65in,angle=-90} \hspace*{-1.37 cm}
\epsfig{file=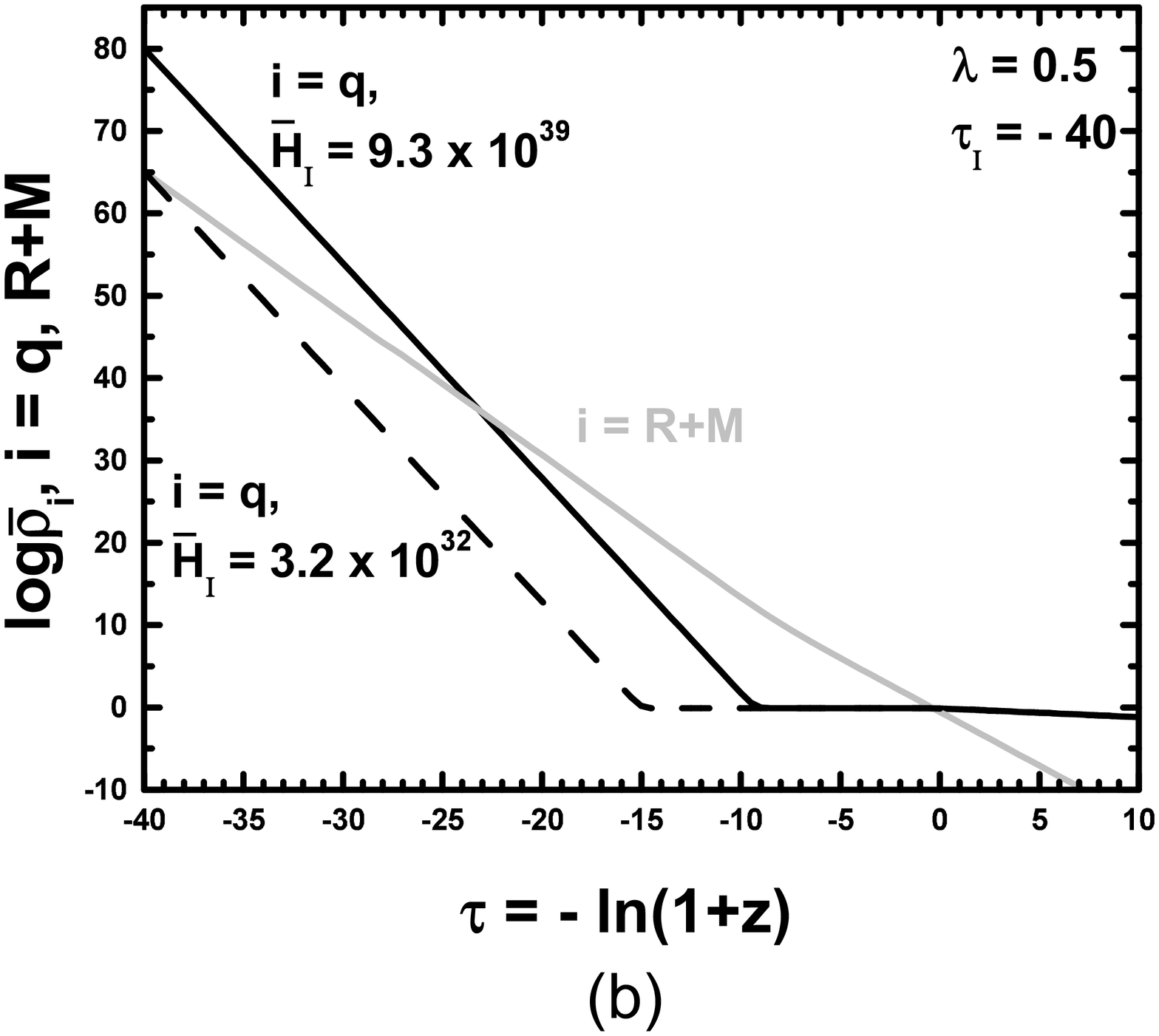,height=3.65in,angle=-90} \hfill
\end{minipage}
\hfill \caption[]{\sl The evolution of the quantities
$\log\vrho_i$ with $i=q$ (black lines) and R+M (light grey line)
as a function of $\vtauf$ for $\lambda=0.5$ and: $\vH_{\rm I
}=10^{40}$ and $\vtif=-40.025~(\vVo=1.6\times10^9)$
[$\vtif=-48.64~(\vVo=2.3)$] (solid [dashed] line) {\sf (a)},
$\vtif=-40$ and $\vHi=9.3\times10^{39}~(\vVo=1.6\times 10^9)$
[$\vH_{\rm I }=3.2\times10^{32}~(\vVo=2.3)$] (solid [dashed] line)
{\sf (b)}.} \label{figr}
\end{figure}

Analyzing comparatively figs. \ref{figr}-{\sf (a)}, {\sf (b)} and
{\sf (c)}, we can demonstrate the characteristic features of the
cosmological history in the presence of $q$. In particular:

\begin{itemize}
\item[{\sf (i)}] For $\vti\leq\vtau\leq\vtkr=-26.2$,
the universe undergoes the KD era. The field $q$ increases
according to eq.~(\ref{qk}) along the left inclined part of the
curve in fig.~\ref{figw}-{\sf (a)} and $\vrho_q$ decreases, more
steeply than $\vrho_{_{\rm R }}$, according to eq.~(\ref{rhok}),
along the left inclined part of the black solid curve in
fig.~\ref{figw}-{\sf (b)}. During this period, $w_q=1$, as shown
in fig.~\ref{figw}-{\sf (c)}. This era terminates at $\vtkr$,
where an intersection of $\vrho_q~[\Omega_q]$ with $\vrho_{_{\rm
R}}~[\Omega_{\rm R}]$ is observed in fig.~\ref{figw}-{\sf
(b)}~[fig.~\ref{figw}-{\sf (c)}].

\item[{\sf (ii)}] For $\vtkr=-26.2\leq\vtau\leq\vtau_{_{\rm
FA}}=0.67$, the universe undergoes successively the RD era and
then the MD era until the re-appearance of DE. More precisely, $q$
freezes to its constant value $\vq_{_{\rm F}}\simeq22.4$ according
to eq.~(\ref{qf}) (or eq.~(\ref{qff}) for $\vtau\geq\vtau_{_{\rm
KF}}=-20.2$) along the horizontal part of the curve in
fig.~\ref{figw}-{\sf (a)}. $\Omega_{\rm R}>\Omega_q$ increases
towards 1 along the light gray line of fig.~\ref{figw}-{\sf (c)}
and then decreases until $\Omega_{_{\rm M}}=\Omega_{_{\rm R}}$ at
$\vtau_{\rm eq}=-8.16$, where a slight kink is observed on the
light gray line of fig.~\ref{figw}-{\sf (b)}. For $\vtau_{_{\rm
KF}}\leq\vtau\leq\vtau_{_{\rm PL}}=-10.22$, $\vrho_q<\vrho_{_{\rm
R}}$ continues to decrease steeply according to eq.~(\ref{rhok}),
along the left inclined part of the black solid curve in
fig.~\ref{figw}-{\sf (b)}, while $w_q$ continues to be 1 as shown
in fig.~\ref{figw}-{\sf (c)}. On the other hand, for $\vtau_{_{\rm
PL}}\leq\vtau\leq\vtau_{_{\rm FA}}$, $\vrho_q$ freezes to its
constant value $\vrho_{q_{\rm F}}=\vV(\vq_{_{\rm F}})=0.62$ in
fig.~\ref{figw}-{\sf (b)}, while $w_q$ transits from 1 to -1 as
shown in fig.~\ref{figw}-{\sf (c)}. At present, we obtain
$w_q(0)=-0.96$, within the limits of eq.~(\ref{wq0}).

\item[{\sf (iii)}] For $\vtau\geq\vtau_{_{\rm FA}}=0.67$,
the universe undergoes a $q$-dominated phase. The field $q$
increases according to eq.~(\ref{qA}) along the right inclined
part of the curve in fig.~\ref{figw}-{\sf (a)}, $\vrho_q$
decreases according to eq.~(\ref{attr}) along the right inclined
parts of the curves in fig.~\ref{figw}-{\sf (b)} for $\lambda=0.1$
and 1.15, while $w_q$ in fig.~\ref{figw}-{\sf (c)} tends to its
fixed-point value for $\lambda=0.5$, $w_q^{\rm fp}=-0.92$. As
shown in the same figure $\Omega^{\rm fp}_q=1$, which identifies
the late-time attractor.
\end{itemize}
Note that, contrary to the case with $\lambda>2$ \cite{prof}, the
variation of $\lambda$ does not affect essentially the position of
the FD plateau but only changes the inclination of the AD curve.

The dependence of the $\vrho_q$ evolution on $\vti~[\vHi]$, can be
easily concluded from fig.~\ref{figr}-{\sf (a)}
[fig.~\ref{figr}-{\sf (b)}], where we plot for $\lambda=0.5$,
$\log\vrho_q$ and $\log\vrho_{_{\rm R+M}}$ versus $\vtau$. In
fig.~\ref{figr}-{\sf (a)} [fig.~\ref{figr}-{\sf (b)}] we take
$\vH_{\rm I}=10^{40}$ and $\vti=-40.025~(\vVo=1.6\times10^9)$
(solid line) or $\vti=-48.64~(\vVo=2.3)$ (dashed line) [$\vti=-40$
and $\vHi=9.3\times10^{39}~(\vVo=1.6\times 10^9)$ (solid line) or
$\vH_{\rm I }=3.2\times10^{32}~(\vVo=2.3)$ (dashed line)]. It is
obvious that increasing \vti\ or $\vHi$, the left, black inclined
line of the $q$-KD regime moves to the right and consequently,
both $\Omega_q(\vti)$ and $\Omega_q(\vtns)$ increase. So, an upper
[lower] bound on $\vti$ or $\vHi$ can be extracted from
eq.~(\ref{nuc}) [eq.~(\ref{domk})] (see fig. \ref{ltH}). The
saturation of these inequalities is the origin of the chosen lower
[upper] $\vti$ or $\vHi$ in fig.~\ref{figr}.

\subsection{I{\ssz MPOSING THE}
Q{\ssz UINTESSENTIAL} R{\ssz EQUIREMENTS}} \label{Qpar}

\begin{figure}[t]
\hspace*{-.25in}
\begin{minipage}{8in}
\epsfig{file=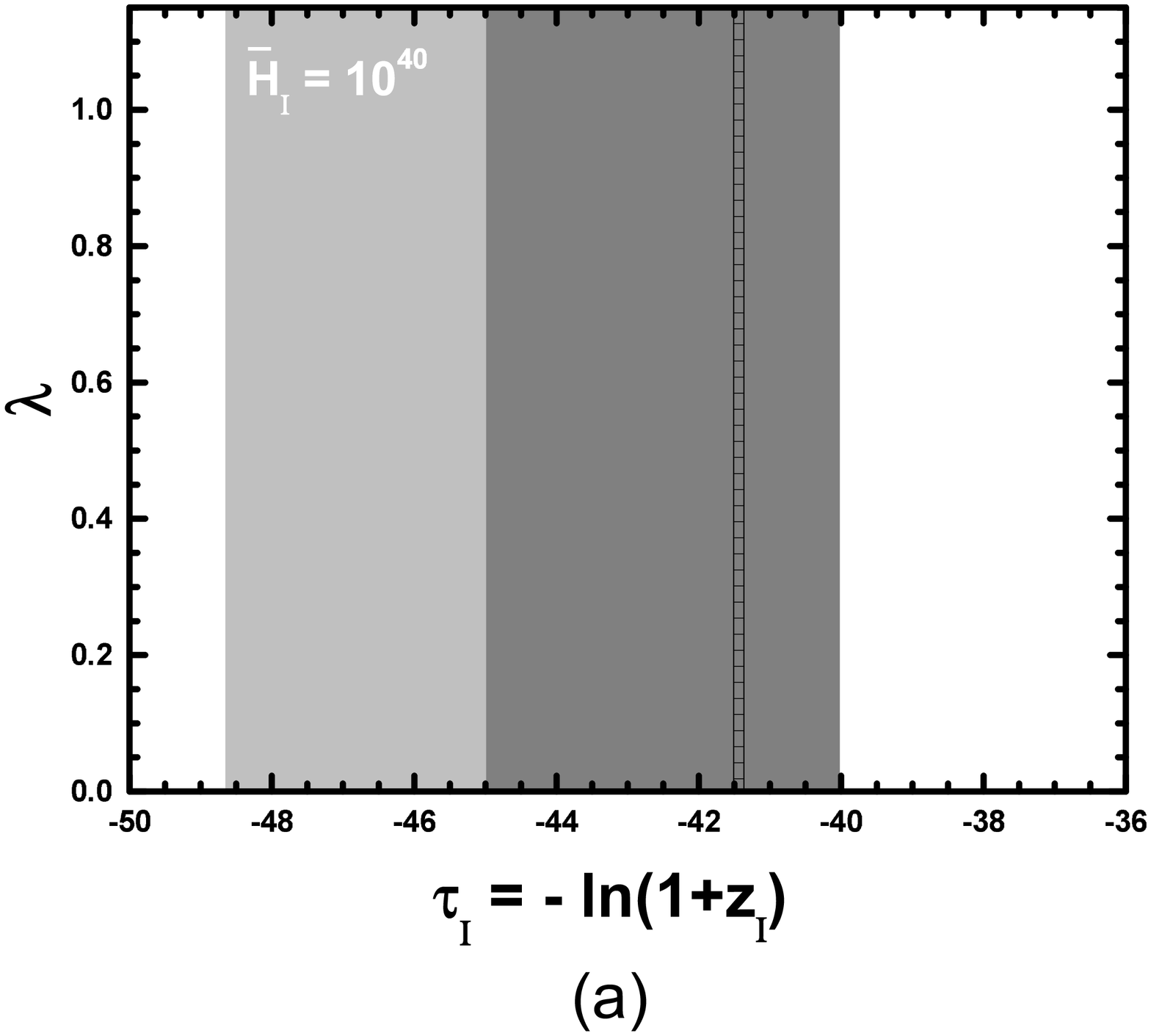,height=3.65in,angle=-90} \hspace*{-1.37 cm}
\epsfig{file=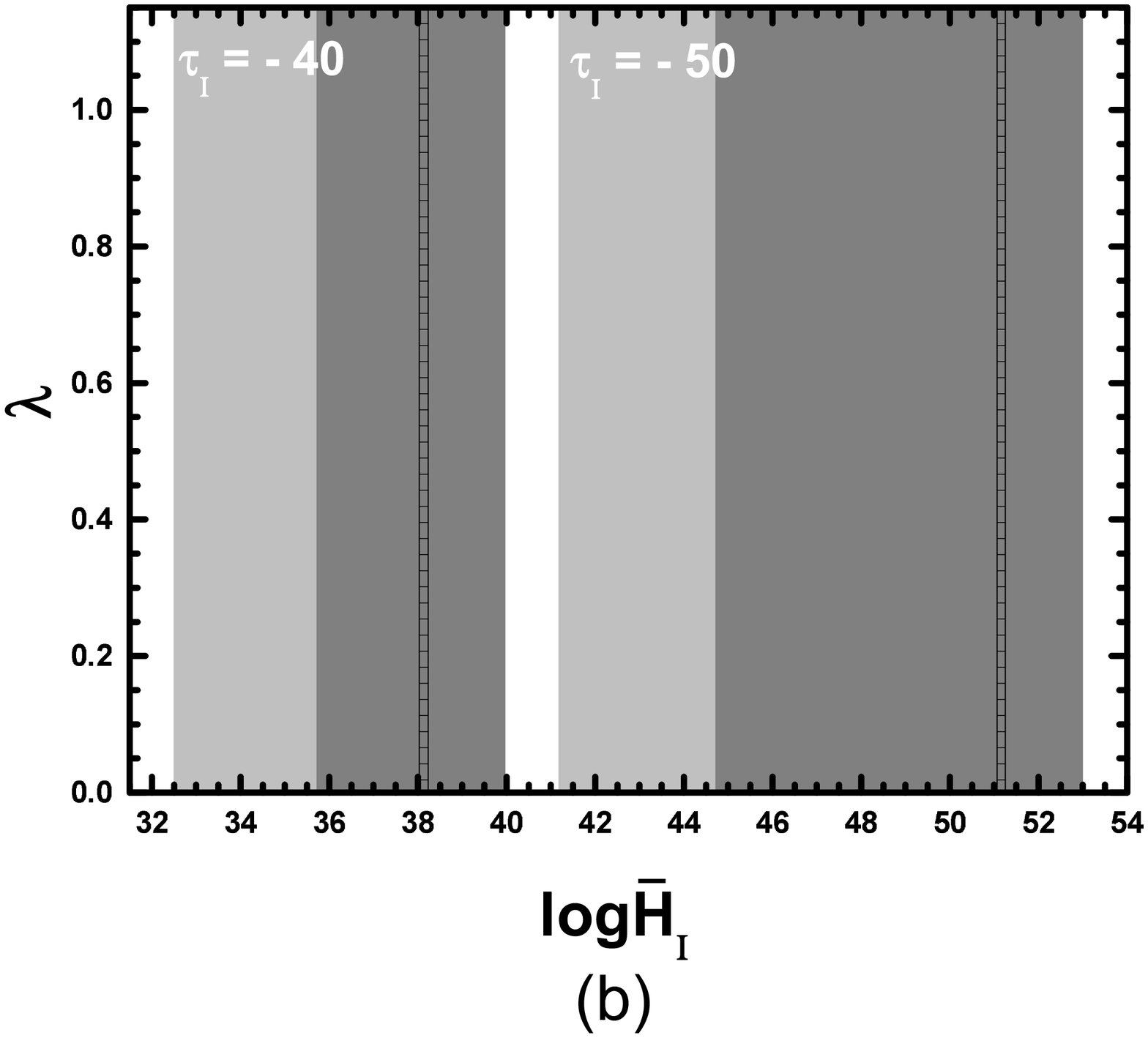,height=3.65in,angle=-90} \hfill
\end{minipage}\vspace*{-.01in}
\hfill\begin{center}
\epsfig{file=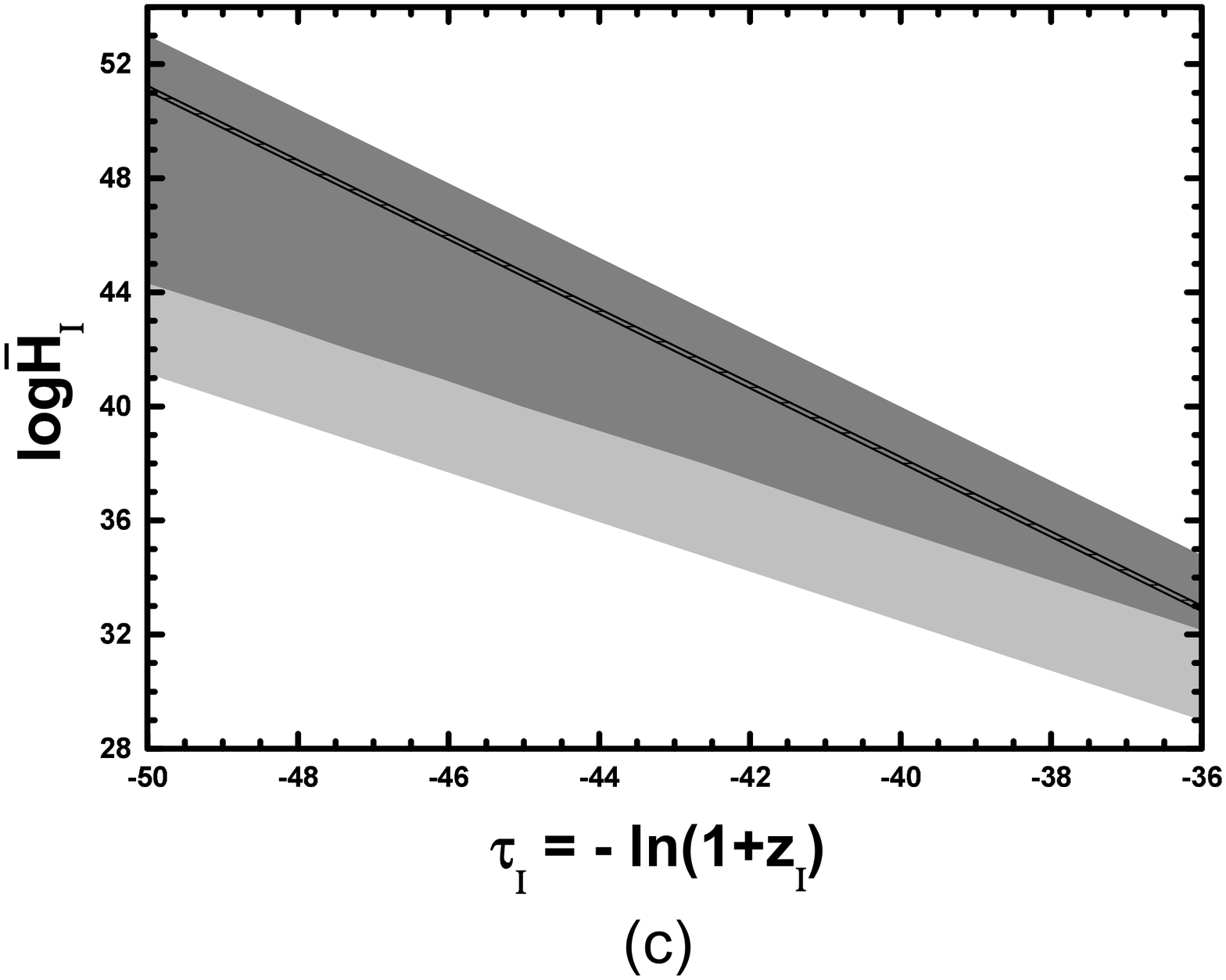,height=3.65in,angle=-90}
\end{center}
\hfill \caption[]{\sl The allowed (shaded) areas on the
$\vtif-\lambda$ plane for $\vHi=10^{40}$ {\sf (a)},
$\log\vHi-\lambda$ plane for $\vtif=-40$ or $\vtif=-50$ {\sf (b)}
and $\vtif-\log\vHi$ plane for $\lambda=0.5$ {\sf (c)}. Lined are,
also, the areas allowed by eq.~(\ref{cdmb}) for $m_\chi=350~{\rm
GeV}$ and $\langle \sigma v \rangle=10^{-7}~{\rm GeV}^{-2}$.}
\label{ltH}
\end{figure}

\hspace{.67cm} We proceed, now in the delineation of the parameter
space of our quintessential model. Agreement with eq.~(\ref{wq0})
entails $0<\lambda\lesssim1.15$ (see also ref.~\cite{german},
where less restrictive upper bound on $w_q(0)$ was imposed). This
range is independent on $\vti$ and $\vHi$ as is shown in
figs.~\ref{ltH}-{\sf (a)} and \ref{ltH}-{\sf (b)}. In these, we
depict respectively the allowed (shaded) regions on the
$\vti-\lambda$ plane for $\vHi=10^{40}$ and on the
$\log\vHi-\lambda$ plane for $\vti=-40$ or $\vti=-50$. In
fig.~\ref{ltH}-{\sf (c)}, we design the allowed area on the
$\vti-\log\vHi$ plane. Although this plot is constructed for
$\lambda=0.5$,  it is obviously $\lambda$ independent. The dark
[light] shaded areas fulfill eq.~(\ref{domk}{\sf
a})~[eq.~(\ref{domk}{\sf b}) and (\ref{domk}{\sf c})]. The right
[left] boundaries of the allowed regions in figs.~\ref{ltH}-{\sf
(a)} and \ref{ltH}-{\sf (b)} are derived from
eq.~(\ref{nuc})~[eq.~(\ref{domk}{\sf b})]. The same origin has the
upper [lower] boundary of the allowed region in
fig.~\ref{ltH}-{\sf (c)}, whereas the left and right boundaries
come from eq.~(\ref{para}{\sf a}). So, for a reasonable set of
($\lambda, \vti, \vHi$), the exponential quintessential model can
become consistent with the observational data \cite{brazil,
german}. The construction of the ruled areas is explained in
sec.~\ref{qparf}.


\subsection{T{\ssz HE} $\Omega_\chi h_0^2$ E{\ssz NHANCEMENT}} \label{omenh}

\hspace{.67cm} The investigation of the $\Omega_\chi h_0^2$
enhancement is the aim of this section. In sec.~\ref{prevpost}, we
illustrate the $\chi$ decoupling during the KD epoch and in
sec.~\ref{omqdep} we examine the dependence of the
$\Omega_{\chi}h_0^2$ increase on $\Omega_q(\vtns)$. Finally, in
sec.~\ref{numan}, we compare the results of our numerical and
semi-analytical $\Omega_{\chi}h_0^2$ calculations.

\begin{figure}[t]
\hspace*{-.25in}
\begin{minipage}{8in}
\epsfig{file=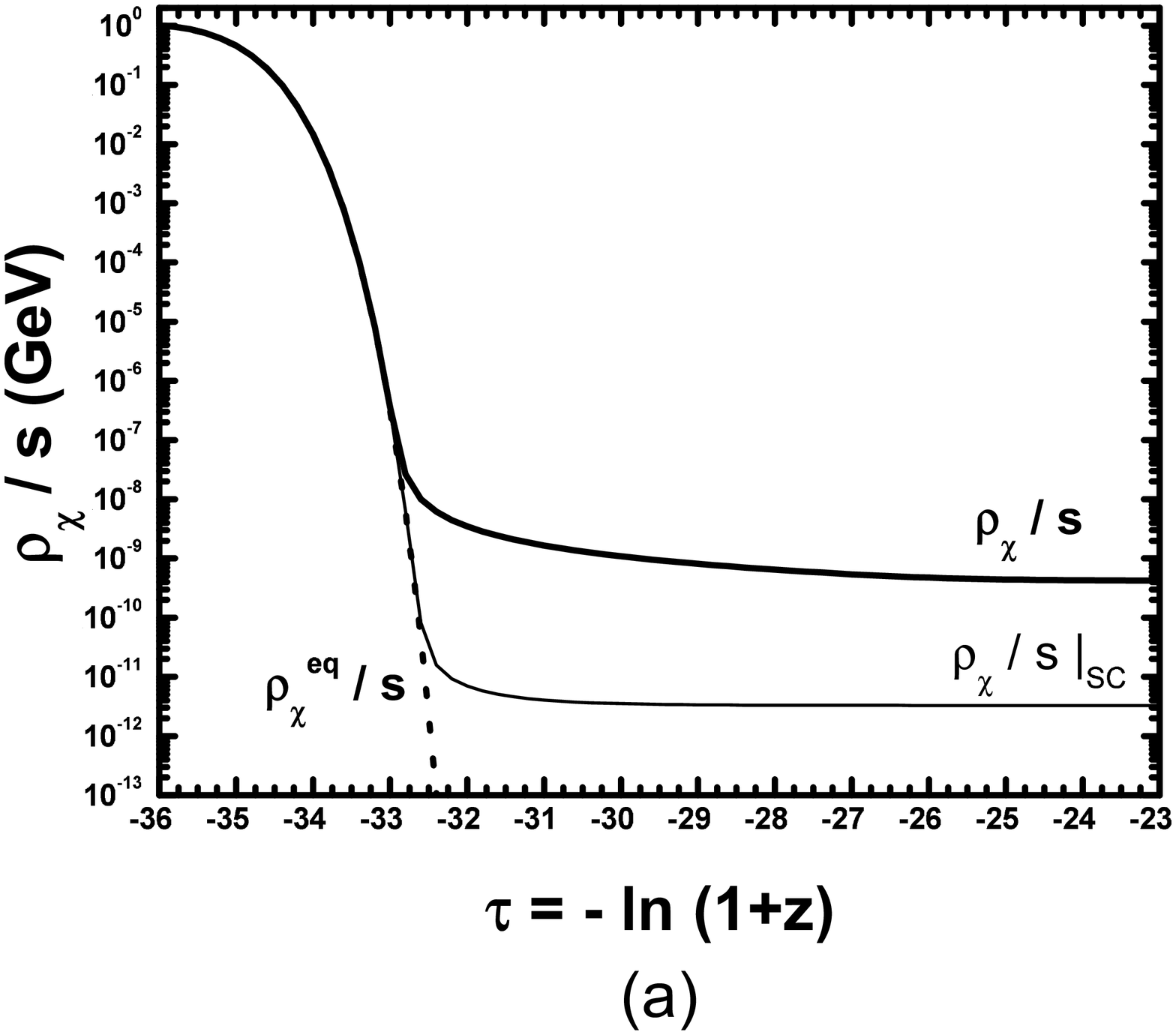,height=3.65in,angle=-90} \hspace*{-1.37 cm}
\epsfig{file=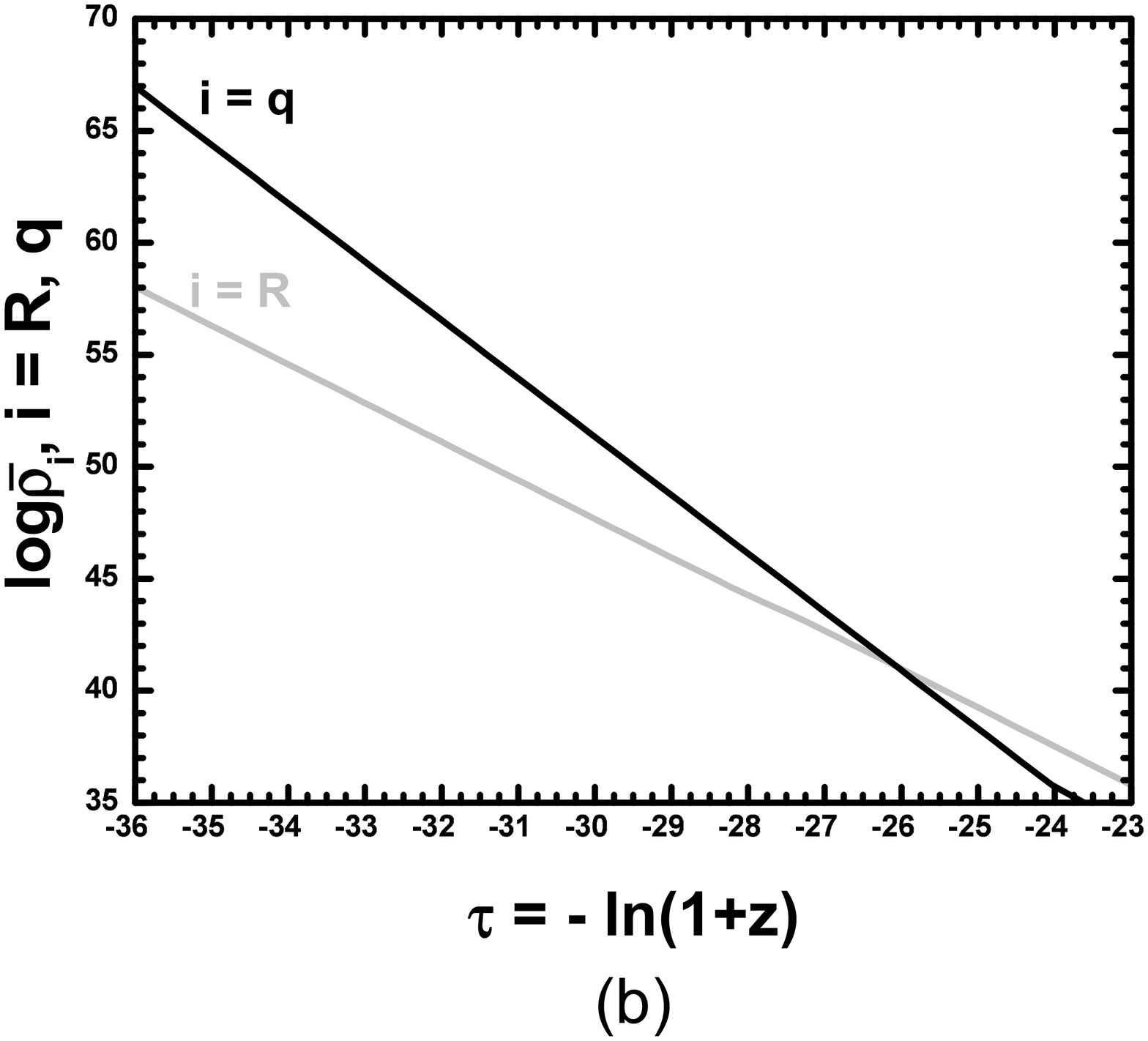,height=3.65in,angle=-90} \hfill
\end{minipage}
\renewcommand{\arraystretch}{1.1}
\begin{center} \begin{tabular}{|c|c|c|c|c|c|} \hline
\multicolumn{6}{|c|}{ \small\sl\bfseries Input Parameters}\\
\hline

$\lambda$&$\vti$& $\vHi$ &$\vVo$ &$m_\chi({\rm GeV})$&$\sv ({\rm
GeV}^{-2})$\\ \hline
$0.5$&{$-41$}
&$10^{40}$&$1.6\times10^8$&$350$&$2.92\times10^{-7}$\\ \hline

\multicolumn{6}{|c|}{\sl\small\bfseries Output Parameters}\\
\hline

$\Omega_q(\vtau_{_{\rm NS}})$&$\vtau_{_{\rm F}}$&$\vtf|_{_{\rm
SC}}$&$\vtau_{_{\rm KR}}$&$\Omega_\chi
h_0^2$&$\Delta\Omega_\chi$\\ \hline

$0.0007$&$-32.9$&$-32.6$&$-25.7$&{$0.115$}&$126.8$
\\ \hline
\end{tabular}\end{center}\renewcommand{\arraystretch}{1.0}

\hfill \caption[]{\sl The evolution as a function of $\vtauf$ of
the quantities $\rho^{\rm eq}_\chi/s$ (dotted line) and
$\rho_\chi/s$ (bold [thin] solid lines) for the quintessential
scenario [SC] ${\sf (a)}$ and $\log\vrho_i$ with $i=q~[i=R]$
(black [light grey] line) ${\sf (b)}$ for the input quantities
listed in the table above.} \label{prepost}
\end{figure}

%
\subsubsection{The $\chi$ decoupling.} \label{prevpost} The $\chi$
decoupling during the KD era is instructively displayed in
figs.~\ref{prepost}-{\sf (a)} and {\sf (b)}. In
fig.~\ref{prepost}-{\sf (a)} we depict $\rho^{[\rm eq]}_{\chi}/s$
(dotted lines) and $\rho_{\chi}/s~[\rho_{\chi}/s|_{_{\rm
SC}}~(\mbox{for} ~g_q=1)]$ (bold [thin] solid lines) versus
$\vtau$. In fig.~\ref{prepost}-${\sf (b)}$, we plot
$\log\vrho_{q}~[\log\vrho_{_{\rm R}}]$ (black [light gray] line)
versus $\vtau$. The needed for our calculation inputs and some
key-outputs are listed in the relevant table. For better
comparison, we give, also, the point of the $\chi$ decoupling, in
the case of the SC ($g_q=1$), $\vtf|_{_{\rm SC}}$. In the present
case, the $\chi$ decoupling is realized deeply within the KD
regime, $\vti<\vtf<\vtkr$ and $\vtf<\vtf|_{_{\rm SC}}$. By
adjusting $\langle\sigma v\rangle$ we extract the central
$\Omega_{\chi}h_0^2$ in eq.~(\ref{cdmb}). The presence of the KD
era causes an efficient $\Omega_{\chi}h_0^2$ enhancement,
$\Delta\Omega_\chi=126.8$ ($\Omega_{\chi}h_0^2|_{_{\rm
SC}}=0.0009$). Note that the condition $\vtf<\vtkr$ is
indispensable in order to obtain sizable $\Delta\Omega_\chi$. This
can be understood as follows. From eqs.~(\ref{tkr}) and
(\ref{rqns}) we obtain $\vtkr\simeq \vtns+\ln r_{_{\rm NS}}/2$. If
we demand $\vtkr<\vtf$, we obtain $\Omega_q(\vtns) <7.7 \times
10^{-10}$, which causes a very weak $\Delta\Omega_\chi<2.2$.
Finally, the phenomenon of re-annihilation \cite{masiero} is not
observed in this context. This is, because in our case $H$
smoothly evolves from its KD to RD behaviour -- see eq.~(\ref{gq})
-- and does not sharply drop after the $\chi$-decoupling as in the
case of ref.~\cite{masiero}.

\subsubsection{The dependence of $\Delta\Omega_\chi$ on $\Omega_q(\vtns)$.}
\label{omqdep} 
\begin{figure}[!t]
\hspace*{-.25in}
\begin{minipage}{8in}
\epsfig{file=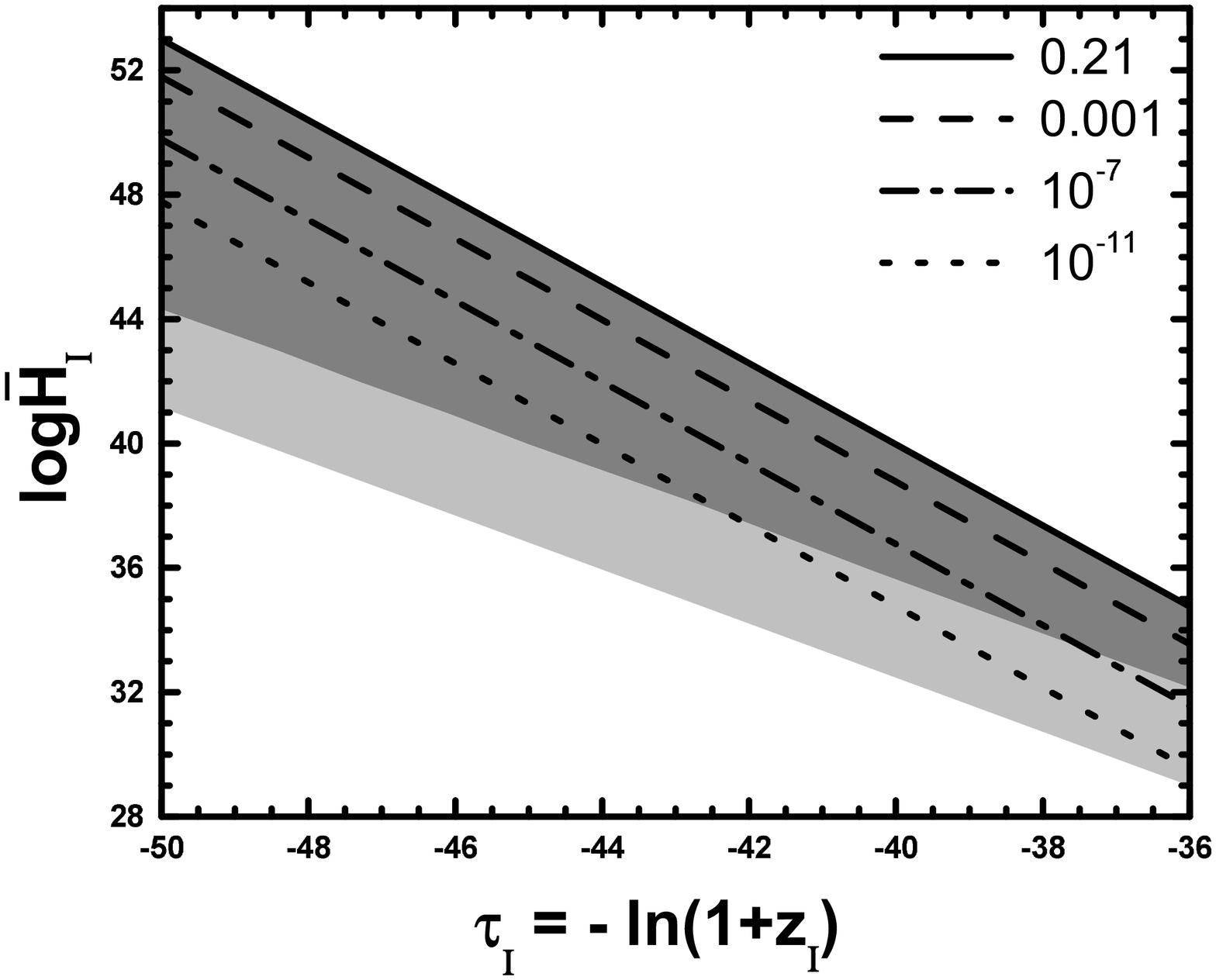,height=3.65in,angle=-90} \hspace*{2.2in}
\epsfig{file=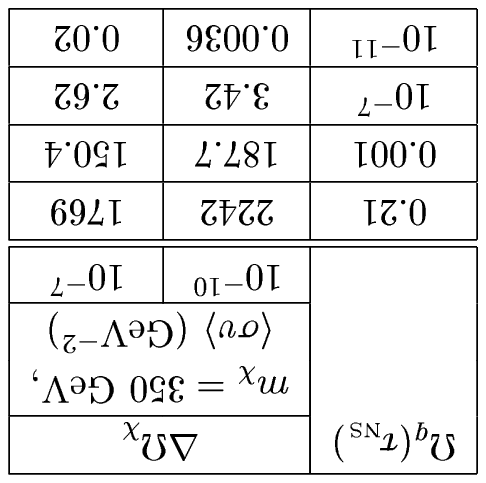,height=2.2in,angle=180} \hfill
\end{minipage}
\hfill \caption[]{\sl Lines with constant $\Omega_q(\vtauf_{_{\rm
NS}})$, indicated in the plot, in the allowed $\vtif-\log\vHi$
plane. The corresponding constant values of $\Delta\Omega_{\chi}$
for $m_{\chi}=350~{\rm GeV}$  and $\langle\sigma
v\rangle=10^{-10}~{\rm GeV^{-2}}$ or $\langle\sigma
v\rangle=10^{-7}~{\rm GeV^{-2}}$ are also listed in the
table.}\label{Htq}
\end{figure}

As is shown in figs.~\ref{figr}-{\sf (a)} and
\ref{figr}-{\sf (b)}, the position of the inclined left part of
the black line (corresponding to $\log\vrho_q$) is affected
crucially by a possible variation of $\vti$ or $\vHi$ but not of
$\lambda$. Therefore, $\Omega_{q}(\vtns)$ and consequently,
$\Delta\Omega_{\chi}$ (see eqs.~(\ref{gq}) and (\ref{rqns}))
depend on $\vti$ or $\vHi$ but not on $\lambda$ (contrary to the
case of ref.~\cite{prof}). Moreover, the dependence of
$\Delta\Omega_\chi$ on $\vti$ or $\vHi$ can be expressed
exclusively as a single-valued function of $\Omega_q(\vtns)$,
since only $g_q$ is involved in the $\Omega_\chi h_0^2$
calculation (see eqs.~(\ref{om1}) and (\ref{BEsol})). This is
illustrated in the fig.~\ref{Htq}, where we depict
iso-$\Omega_q(\vtns)$ lines on the allowed region of the
$\vti-\log\vHi$ plane, presented in fig.~\ref{ltH}-{(c)}. Along
these lines, $\Delta\Omega_{\chi}$ remains, also, constant
(indicated in the table of fig.~\ref{Htq}) for fixed
$m_{\chi}=350~{\rm GeV}$ and $\langle\sigma v\rangle=10^{-10}~{\rm
GeV^{-2}}$ or $\langle\sigma v\rangle=10^{-7}~{\rm GeV^{-2}}$.
Consequently, the number of the free $q$ parameters
$(\vti,~\vHi)$, which determine $\Delta\Omega_{\chi}$, can be
reduced by one and replaced by $\Omega_q(\vtns)$.

\subsubsection{Numerical Versus Semi-Analytical Results.}
\label{numan} 
\begin{figure}[!t]\vspace*{-.15in}
\hspace*{-.25in}
\begin{minipage}{8in}
\epsfig{file=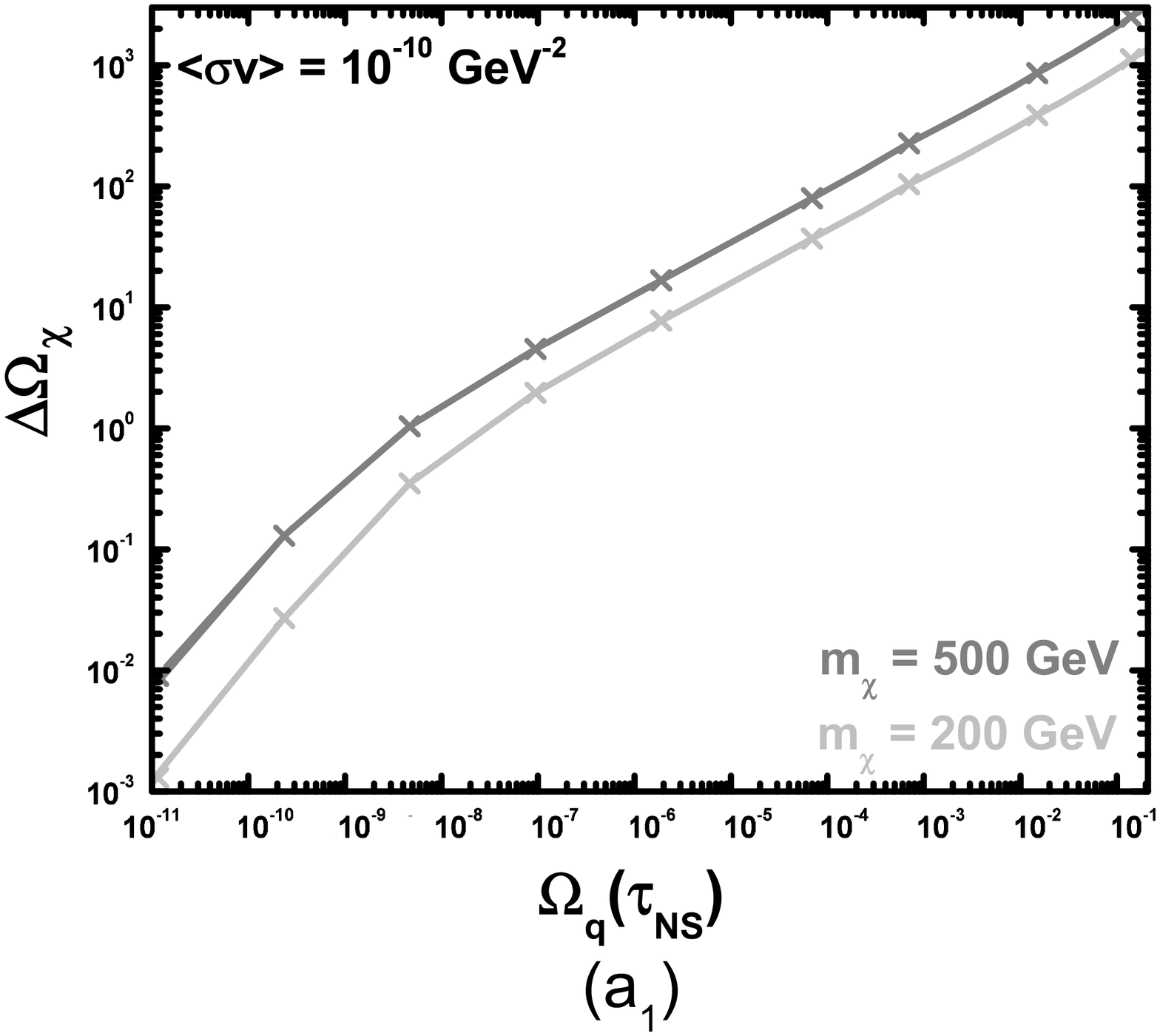,height=3.65in,angle=-90} \hspace*{-1.37 cm}
\epsfig{file=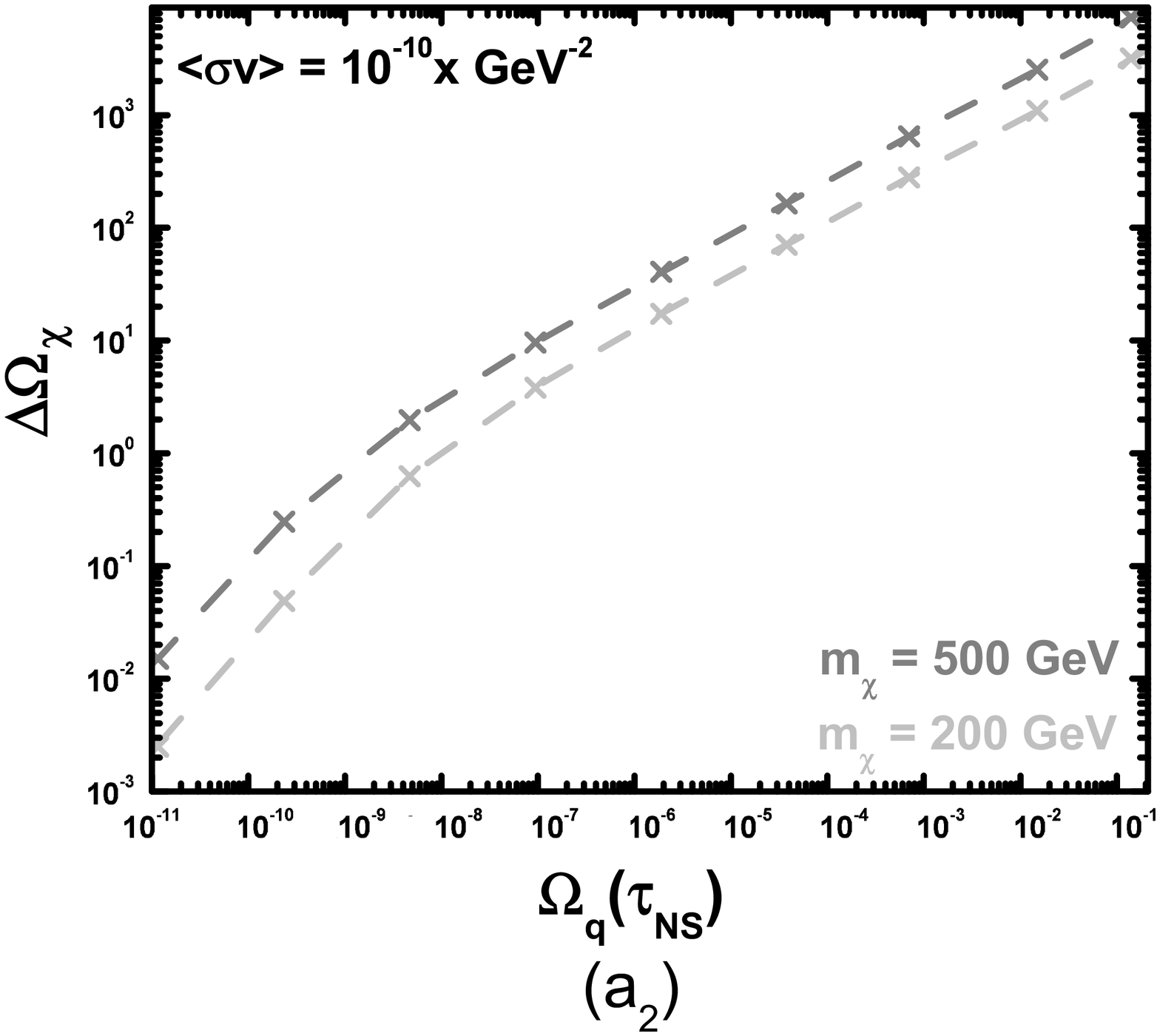,height=3.65in,angle=-90} \hfill
\end{minipage}\vspace*{-.01in}
\hfill\hspace*{-.25in}
\begin{minipage}{8in}
\epsfig{file=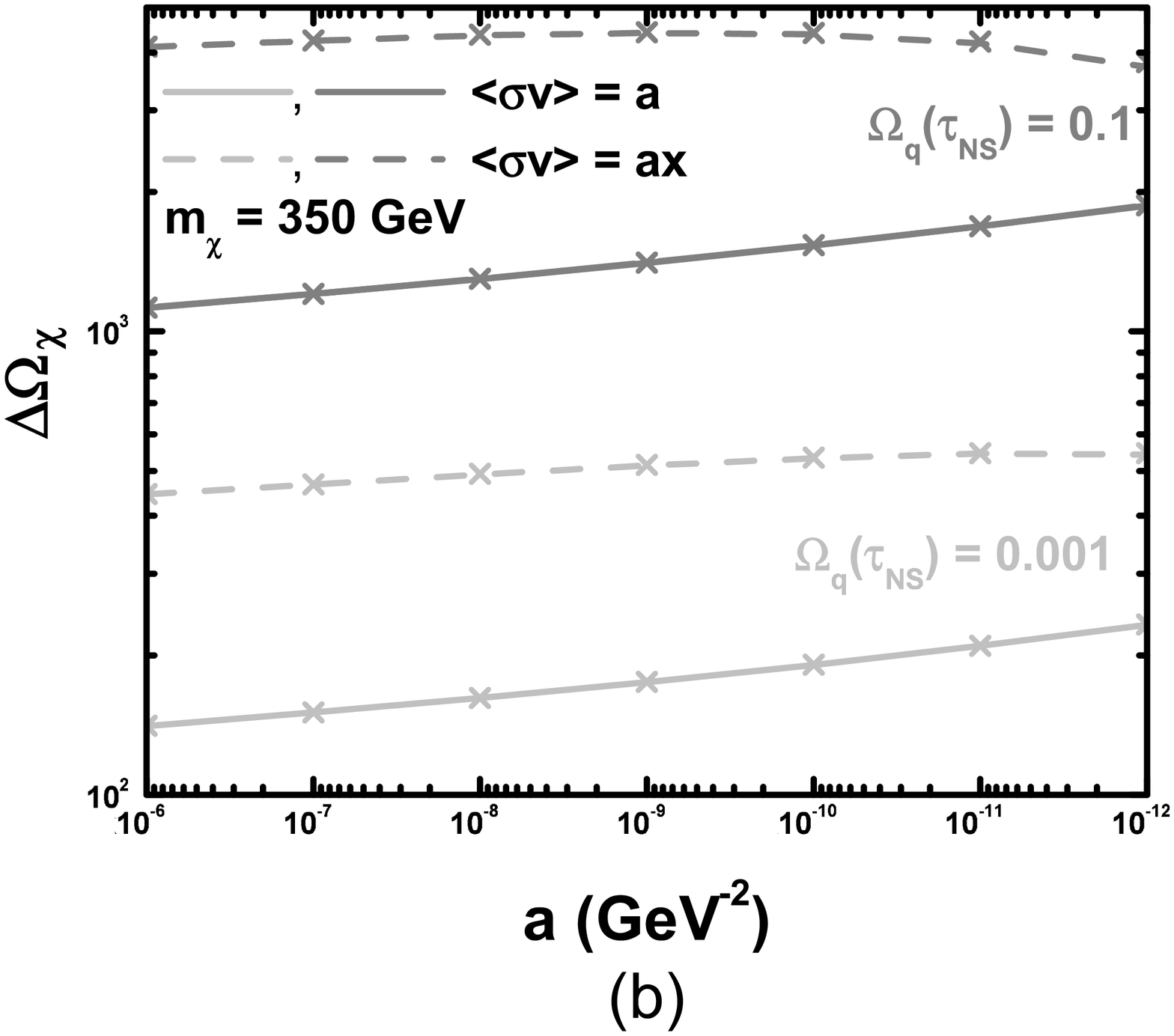,height=3.65in,angle=-90} \hspace*{-1.37 cm}
\epsfig{file=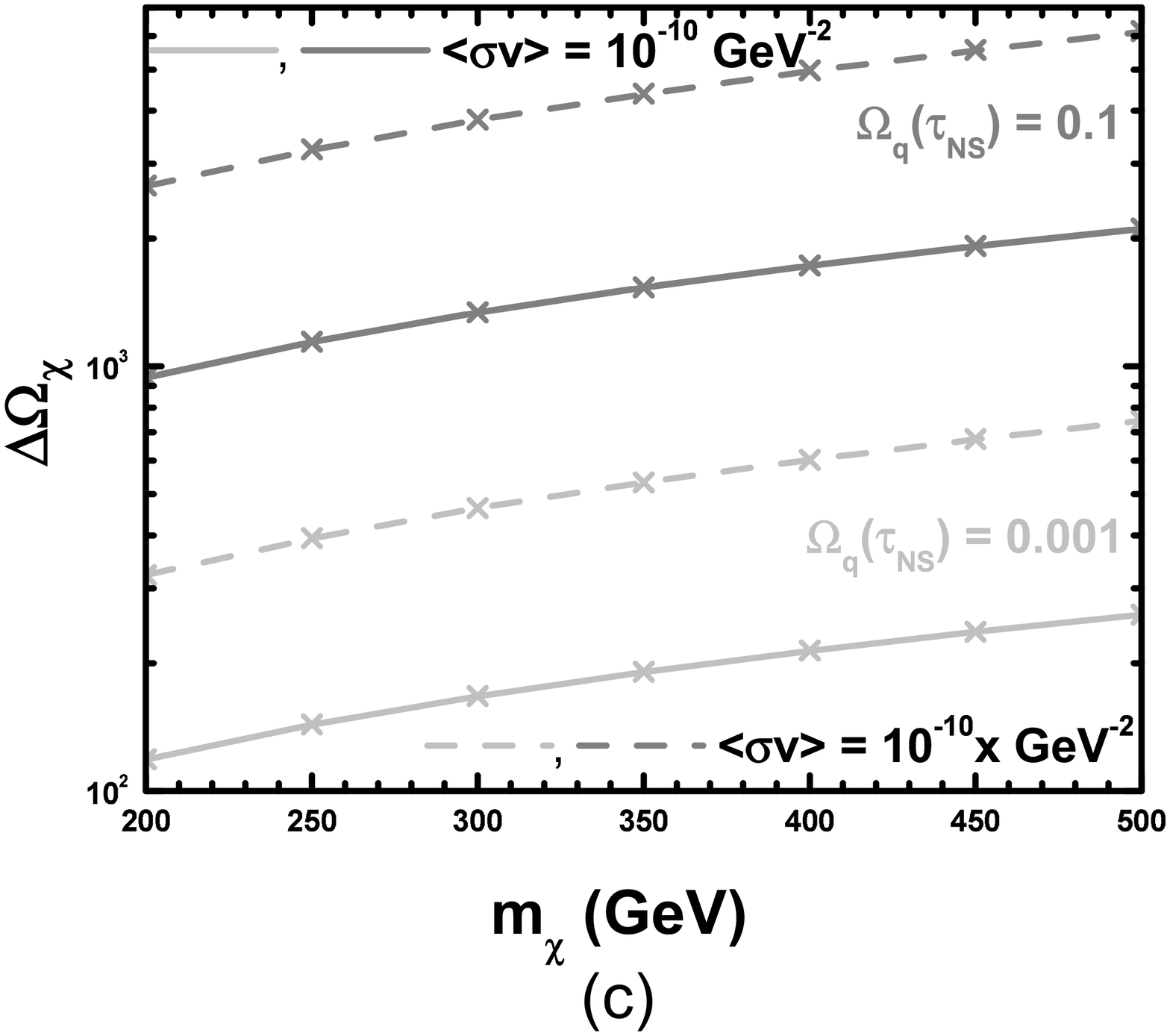,height=3.65in,angle=-90} \hfill
\end{minipage}
\hfill \caption[]{\sl $\Delta\Omega_{\chi}$ versus
$\Omega_q(\vtauf_{_{\rm NS}})$ for $m_{\chi}=200~{\rm
GeV}~[m_{\chi}=500~{\rm GeV}]$ (light [normal] grey lines and
crosses) and $\langle\sigma v\rangle=10^{-10}~{\rm
GeV^{-2}}~[\langle\sigma v\rangle=10^{-10}x~{\rm GeV^{-2}}]~{\sf
(a_1~[a_2])}$. Also, $\Delta\Omega_{\chi}$ versus $a=\langle\sigma
v\rangle~[a=\langle\sigma v\rangle/x]$ for $m_{\chi}=350~{\rm
GeV}$ and $\langle\sigma v\rangle=a~[\langle\sigma v\rangle=ax]$
(solid [dashed] lines) {\sf (b)} and $\Delta\Omega_{\chi}$ versus
$m_\chi$ for $\langle\sigma v\rangle=10^{-10}~{\rm
GeV}^{-2}~[\langle\sigma v\rangle=10^{-10}x~{\rm GeV}^{-2}]$
(solid [dashed] lines) {\sf (c)}. We take $\Omega_q(\vtauf_{_{\rm
NS}})=0.001~[\Omega_q(\vtauf_{_{\rm NS}})=0.1]$ (light [normal]
grey lines and crosses). The solid or dashed lines [crosses] are
obtained by our numerical code [semi-analytical
expressions].}\label{om}
\end{figure}

The validity of our semi-analytical approach can be
tested by comparing its results for $\Delta\Omega_{\chi}$ with
those obtained by the numerical solution of eqs.~(\ref{rx}). In
addition, useful conclusions can be inferred for the behavior of
$\Delta\Omega_{\chi}$ as a function of our free parameters,
$m_\chi,~\sv$ and $\Omega_q(\vtauf_{_{\rm NS}})$. Our results are
presented in fig.~\ref{om}. The solid and dashed lines are drawn
from our numerical code, whereas crosses are obtained by employing
the formulas of sec.~\ref{Seqs} with $\delta_{\rm
F}=1.28~[\delta_{\rm F}=1.35]$ for $\sv=a~[\langle\sigma
v\rangle=bx]$. In figs.~\ref{om}-{$\sf (a_1)$} [\ref{om}-{$\sf
(a_2)$}], we present $\Delta\Omega_{\chi}$ versus
$\Omega_q(\vtau_{_{\rm NS}})$ for $\langle\sigma
v\rangle=10^{-10}~{\rm GeV^{-2}}~[\langle\sigma
v\rangle=10^{-10}x~{\rm GeV^{-2}}]$. We take $m_{\chi}=200~{\rm
GeV}~[m_{\chi}=500~{\rm GeV}]$ (light [normal] grey lines and
crosses). In fig.~\ref{om}-{\sf (b)}, we plot
$\Delta\Omega_{\chi}$ versus $a=\langle\sigma
v\rangle~[a=\langle\sigma v\rangle/x]$ for $m_{\chi}=350~{\rm
GeV}$ and $\langle\sigma v\rangle=a~[\langle\sigma v\rangle=ax]$
(solid [dashed] lines). In fig.~\ref{om}-{\sf (c)}, we depict
$\Delta\Omega_{\chi}$ versus $m_\chi$ for $\langle\sigma
v\rangle=10^{-10}~{\rm GeV}^{-2}~[\langle\sigma
v\rangle=10^{-10}x~{\rm GeV}^{-2}]$ (solid [dashed] lines). In the
last two cases, we use $\Omega_q(\vtau_{_{\rm
NS}})=0.001~[\Omega_q(\vtau_{_{\rm NS}})=0.1]$ (light [normal]
grey lines and crosses). As we anticipated in sec.~\ref{detvar},
$\Delta\Omega_{\chi}$ increases when $\Omega_q(\vtns)$ (see
figs.~\ref{om}-${\sf (a_1)}$ and ${\sf (a_2)}$) or $m_\chi$ (see
fig. ~\ref{om}-{\sf (c)}) increases and when $\sv$ decreases (see
fig.~\ref{om}-{\sf (b)}). From figs.~\ref{om}-{\sf (b)} and
\ref{om}-{\sf (c)} is, also, deduced that $\Delta\Omega_{\chi}$
increases more drastically in the $\sv=bx$ case than in the
$\sv=a$ case for $a=b$ and fixed $m_\chi$ and
$\Omega_q(\vtau_{_{\rm NS}})$. Evident is, finally, the agreement
between numerical and semi-analytical results.

\subsection{I{\ssz MPOSING THE}
CDM R{\ssz EQUIREMENT}} \label{NTR}

\hspace{.565cm} Requiring $\Omega_{\chi}h_0^2$ to be confined in
the cosmologically allowed range of eq.~(\ref{cdmb}), one can
restrict not only the CDM parameters (see subsec.~\ref{svmp}) but
also the $q$ parameters, $\lambda,~\vti$ and $\vHi$ (see
subsec.~\ref{qparf}) or $\Omega_q(\vtns)$ (see subsec.~\ref{omq}).
The data is derived exclusively by the numerical program. Let us
note, in passing, that bounds arisen from eq.~(\ref{cdmb}{\sf b}),
are more rigorous than those originated from eq.~(\ref{cdmb}{\sf
a}), since other production mechanisms of $\chi$'s may be
activated \cite{snr, kohri} and/or other CDM candidates
\cite{candidates} may contribute to $\Omega_{\rm CDM}$.

\begin{figure}[!h]\vspace*{-.15in}
\hspace*{-.25in}
\begin{minipage}{8in}
\epsfig{file=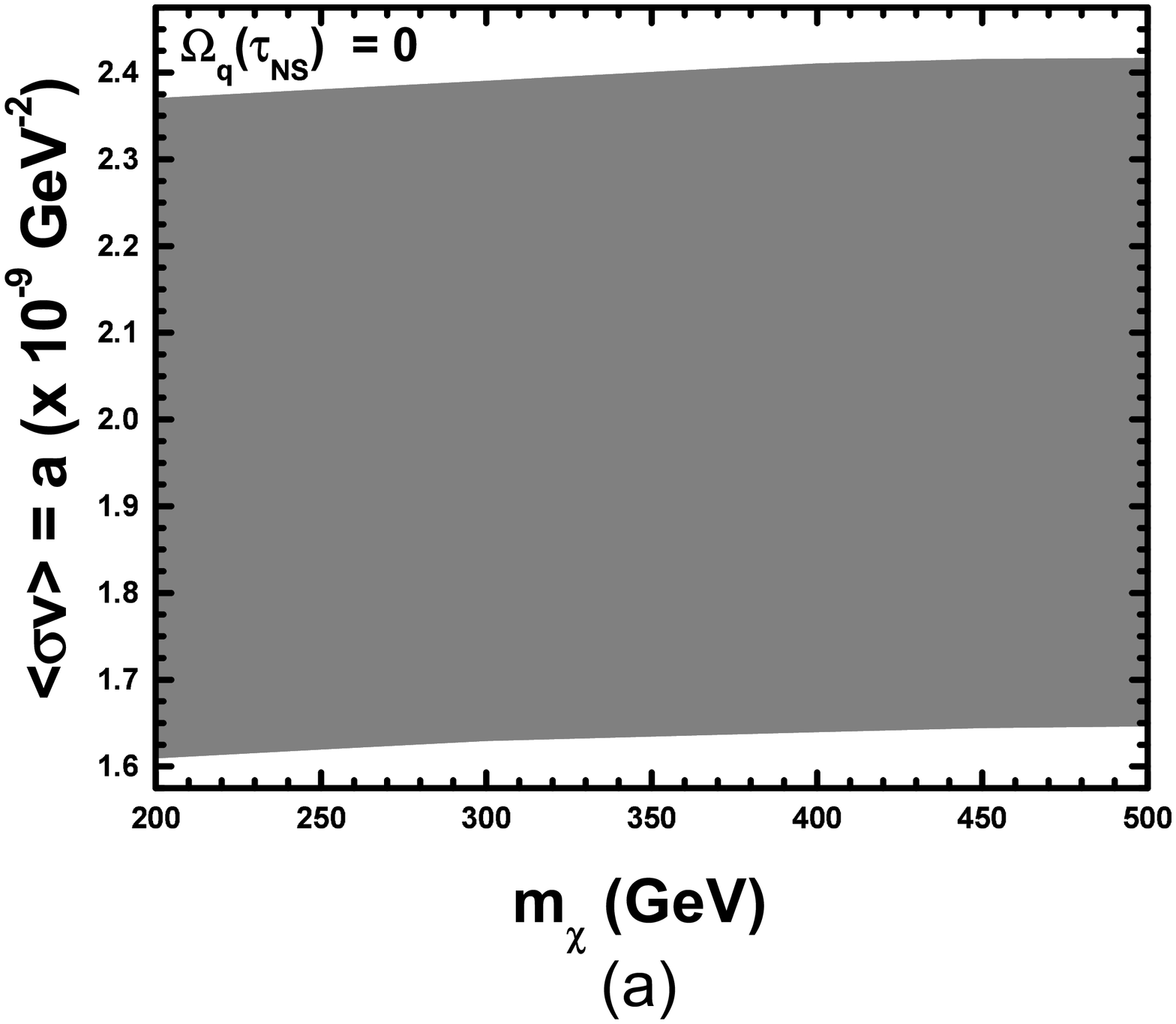,height=3.65in,angle=-90} \hspace*{-1.37 cm}
\epsfig{file=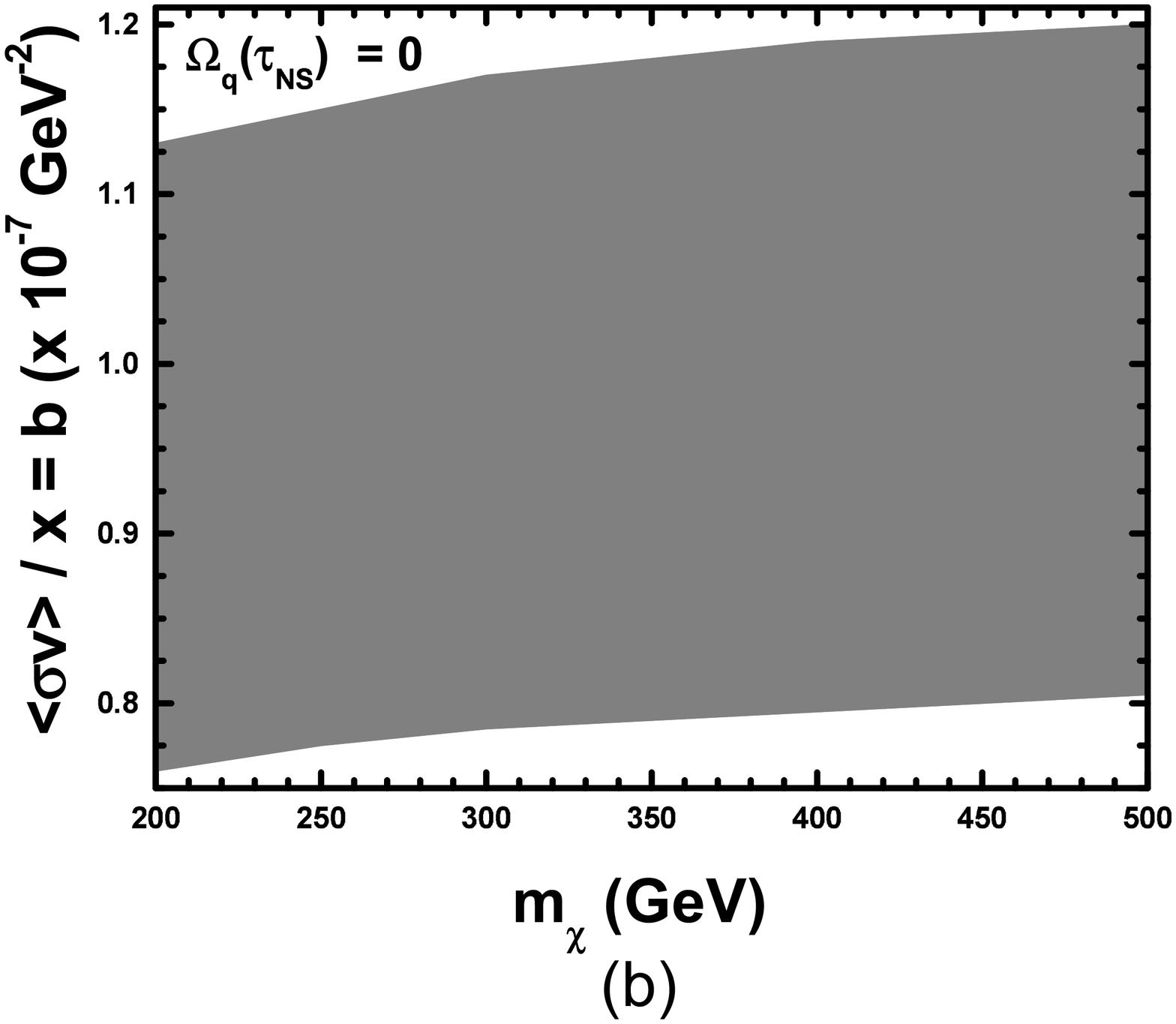,height=3.65in,angle=-90} \hfill
\end{minipage}\vspace*{-.01in}
\hfill\hspace*{-.25in}
\begin{minipage}{8in}
\epsfig{file=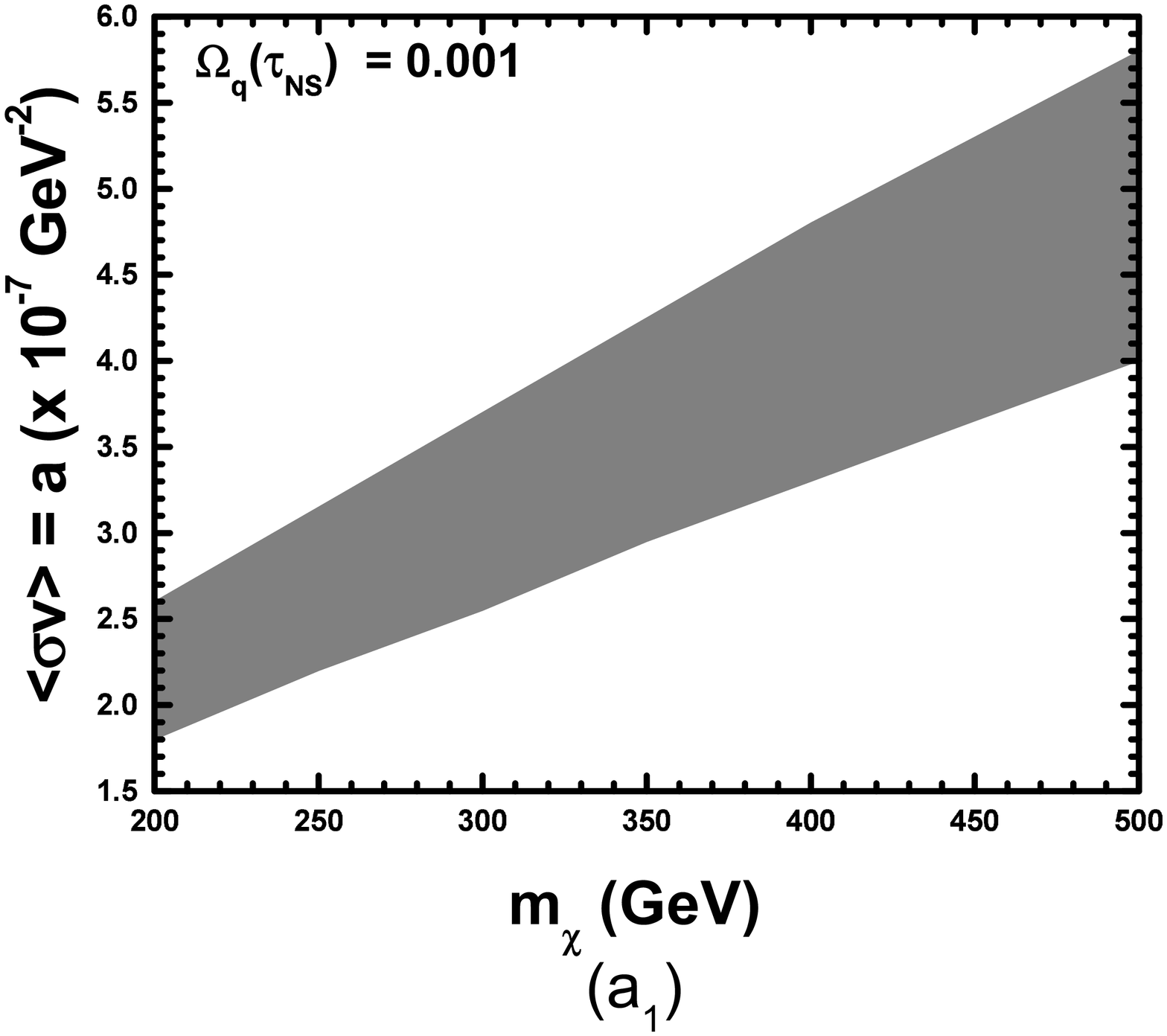,height=3.65in,angle=-90} \hspace*{-1.37 cm}
\epsfig{file=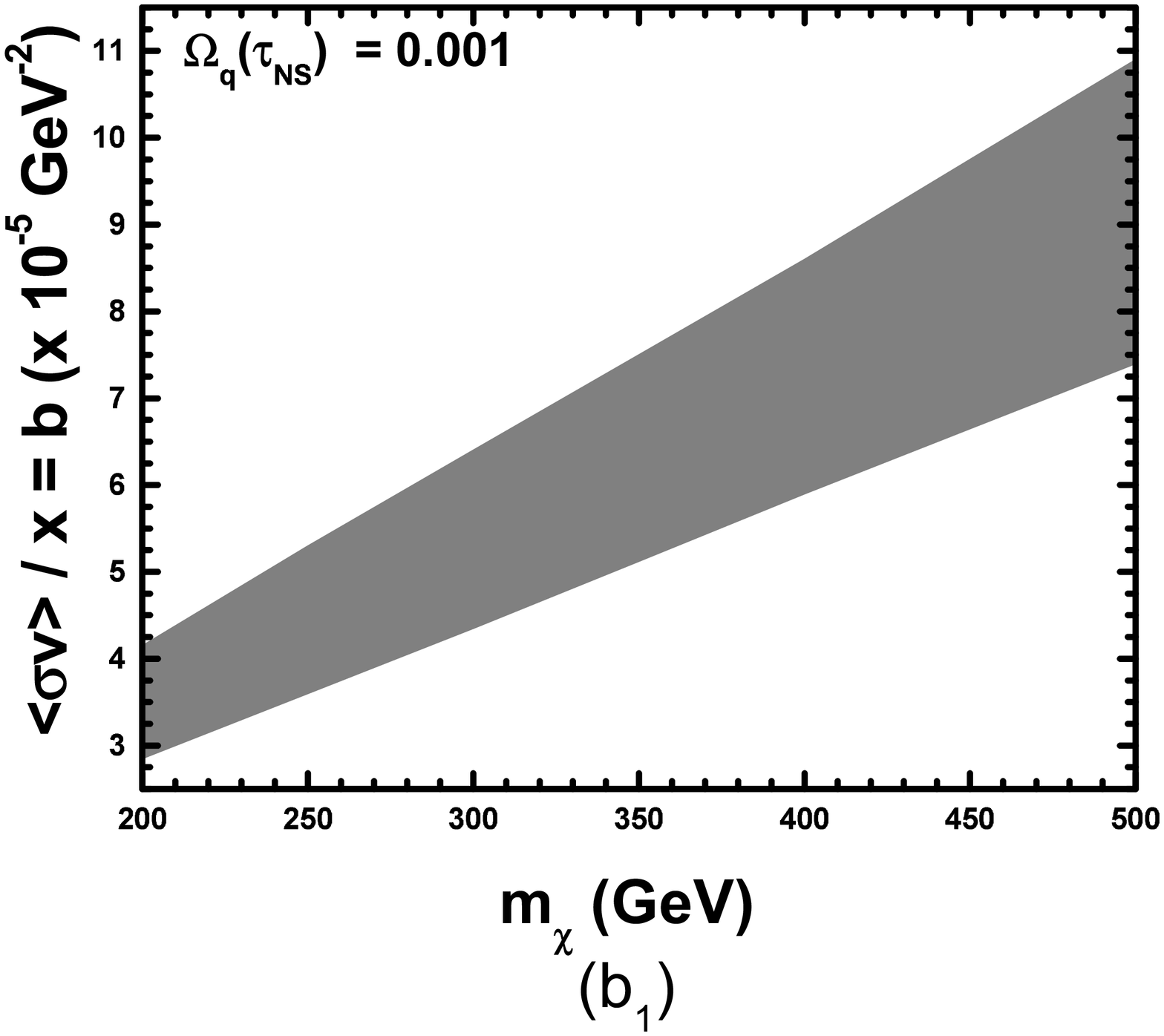,height=3.65in,angle=-90} \hfill
\end{minipage}
\hfill\hspace*{-.25in}
\begin{minipage}{8in}
\epsfig{file=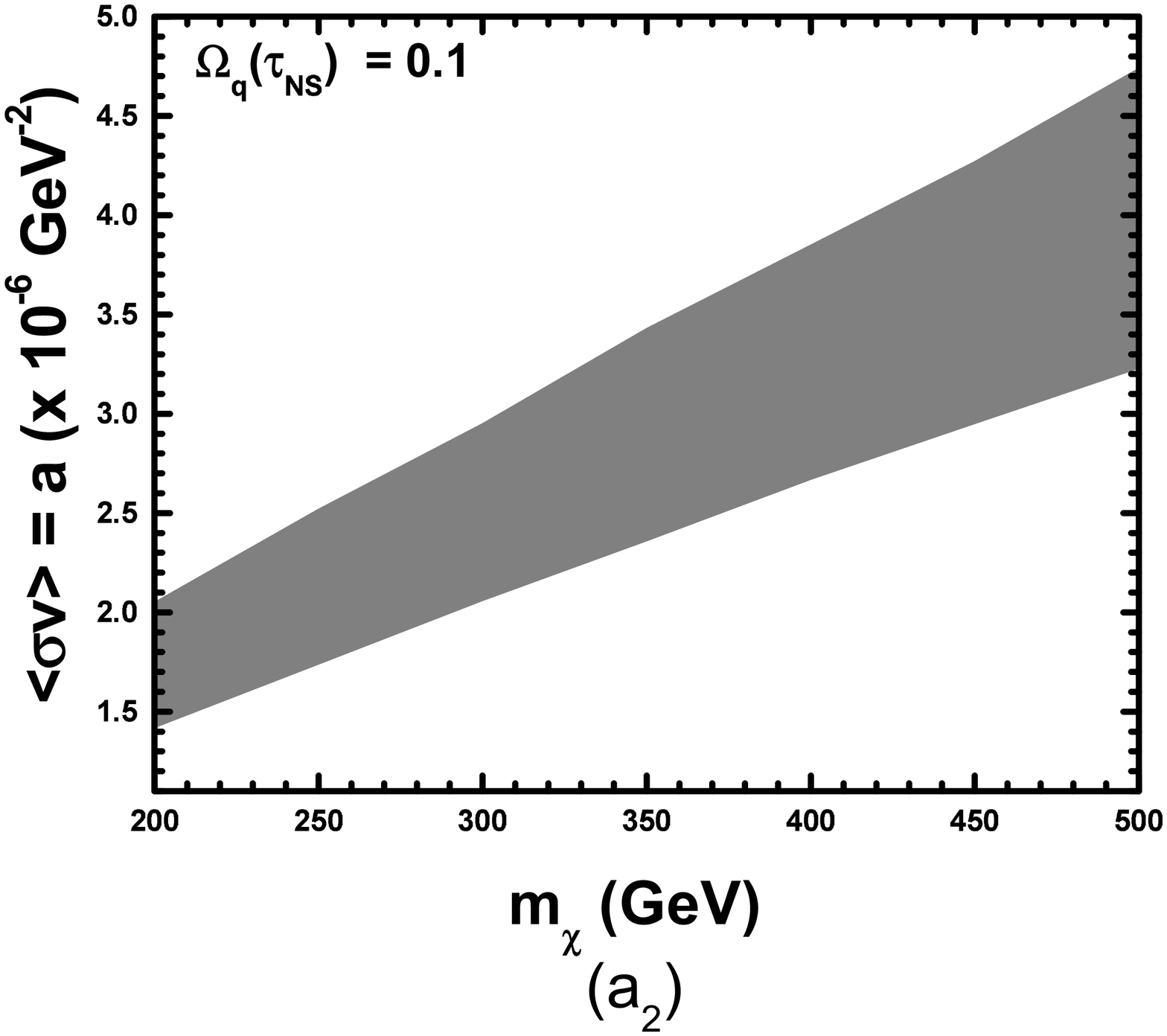,height=3.65in,angle=-90} \hspace*{-1.37 cm}
\epsfig{file=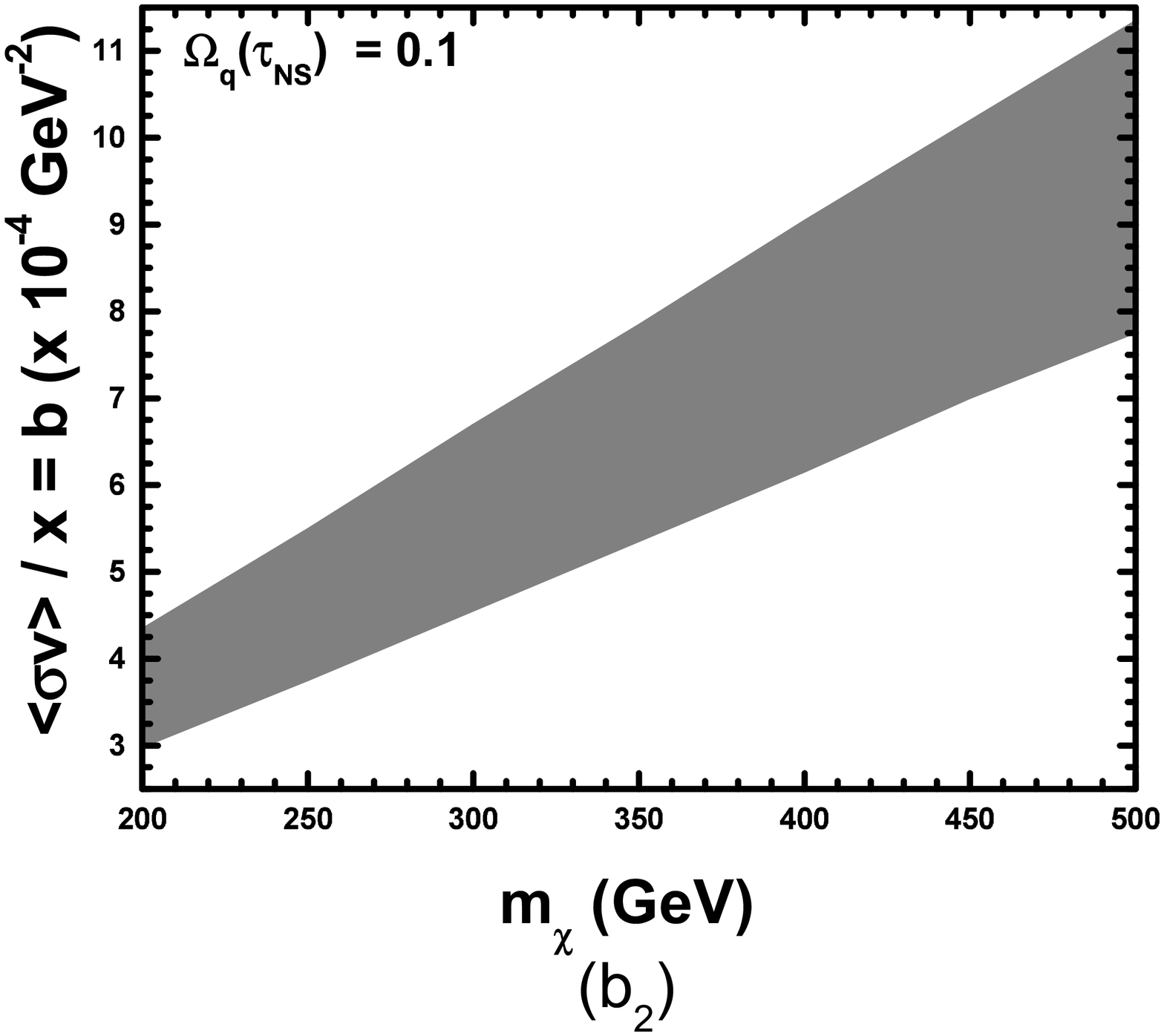,height=3.65in,angle=-90} \hfill
\end{minipage}
\hfill \caption[]{\sl The allowed region on the $m_\chi-\sv$ plane
for $\sv=a$ ${\sf (a,~a_1,~a_2)}$ or on the $m_\chi-\sv/x$ plane
for $\sv=bx$ ${\sf (b,~b_1,~b_2)}$. We take
$\Omega_q(\vtns)=0.001~[\Omega_q(\vtns)=0.1]$ in fig. ${\sf
(a_1,~b_1)}$ [fig. ${\sf (a_2,~b_2)}$] whereas, for the sake of
comparison, we consider the SC ($\Omega_q(\vtns)=0$) in figs. {\sf
(a)} and {\sf (b)}.}\label{svm}
\end{figure}

\subsubsection{Constraining the CDM parameters.}\label{svmp}
Fixing the $q$ parameters, we can derive restrictions on the CDM
parameters. Namely, in fig.~\ref{svm} we construct the allowed
regions on the $m_\chi-\sv$ plane for $\sv=a$ ${\sf
(a,~a_1,~a_2)}$ or on the $m_\chi-\sv/x$ plane  for $\sv=bx$ ${\sf
(b,~b_1,~b_2)}$. As we showed in subsec.~\ref{omqdep}, the $q$
parameters can be replaced by $\Omega_q(\vtns)$. So, in
figs.~\ref{svm}-${\sf (a_1)}$ and ${\sf (b_1)}$
[figs.~\ref{svm}-${\sf (a_2)}$ and ${\sf (b_2)}$], we fix
$\Omega_q(\vtns)=0.001~[\Omega_q(\vtns)=0.1]$, whereas in
figs.~\ref{svm}-{\sf (a)} and {\sf (b)}, we consider, for better
reference, the SC, with $\Omega_q(\vtns)=0$. The upper [lower]
boundaries of the allowed areas are derived from
eq.~(\ref{cdmb}${\sf a}$) [eq.~(\ref{cdmb}${\sf b}$)]. This is due
to the fact that $\Omega_{\chi}h_0^2$ is inverse proportional to
$\sv$ as is obvious from eqs.~(\ref{om1}) and (\ref{BEsol}) and
so, $\Omega_{\chi}h_0^2$ decreases as $\sv$ increases (contrary to
the case of the non-equilibrium $\chi$ production \cite{fornengo,
snr}).

We observe that with
$\Omega_q(\vtns)=0.001~[\Omega_q(\vtns)=0.1]$, agreement with
eq.~(\ref{cdmb}) entails almost two [three] orders of magnitude
higher $\sv$'s than those required in the SC. Also, due to the
presence of $\Omega_q(\vtns)>0$,  $\Omega_\chi h_0^2$ increases
with $m_\chi$ more dramatically than in the case of the SC
($\Omega_q(\vtns)=0$), illustrated  in figs.~\ref{svm}-{\sf (a)}
and {\sf (b)}. This effect is more straightened in the $\sv=bx$
case, as is seen in figs.~\ref{svm}-${\sf (b_1)}$ and ${\sf
(b_2)}$.

The requisite high values for $\sv$ (almost unnatural in the
$\sv=bx$ case) can be obtained by resorting to SUSY models which
ensure $A$-pole effects \cite{lah, quasi} or ``gaugino-inspired''
CANs \cite{edjo, nelson, su5b}, as in the applications \cite{su5,
wells} of ref.~\cite{prof}. Less efficient augmentation of $\sv$
can be achieved by lowering the masses of the CDM candidates in ED
models \cite{lkk, lzk, branon} or  by employing sfermionic CANs
\cite{ellis2, boehm, su5b} in SUSY models. Consequently, the
constrained minimal SUSY model \cite{Cmssm}, although tightly
restricted even in the SC \cite{quasi,spanos}, can become
consistent with a quintessential KD period, e.g. applying the
$A$-pole effects \cite{lah, quasi}.

\begin{figure}[!t]\vspace*{-.15in}
\hspace*{-.25in}
\begin{minipage}{8in}
\epsfig{file=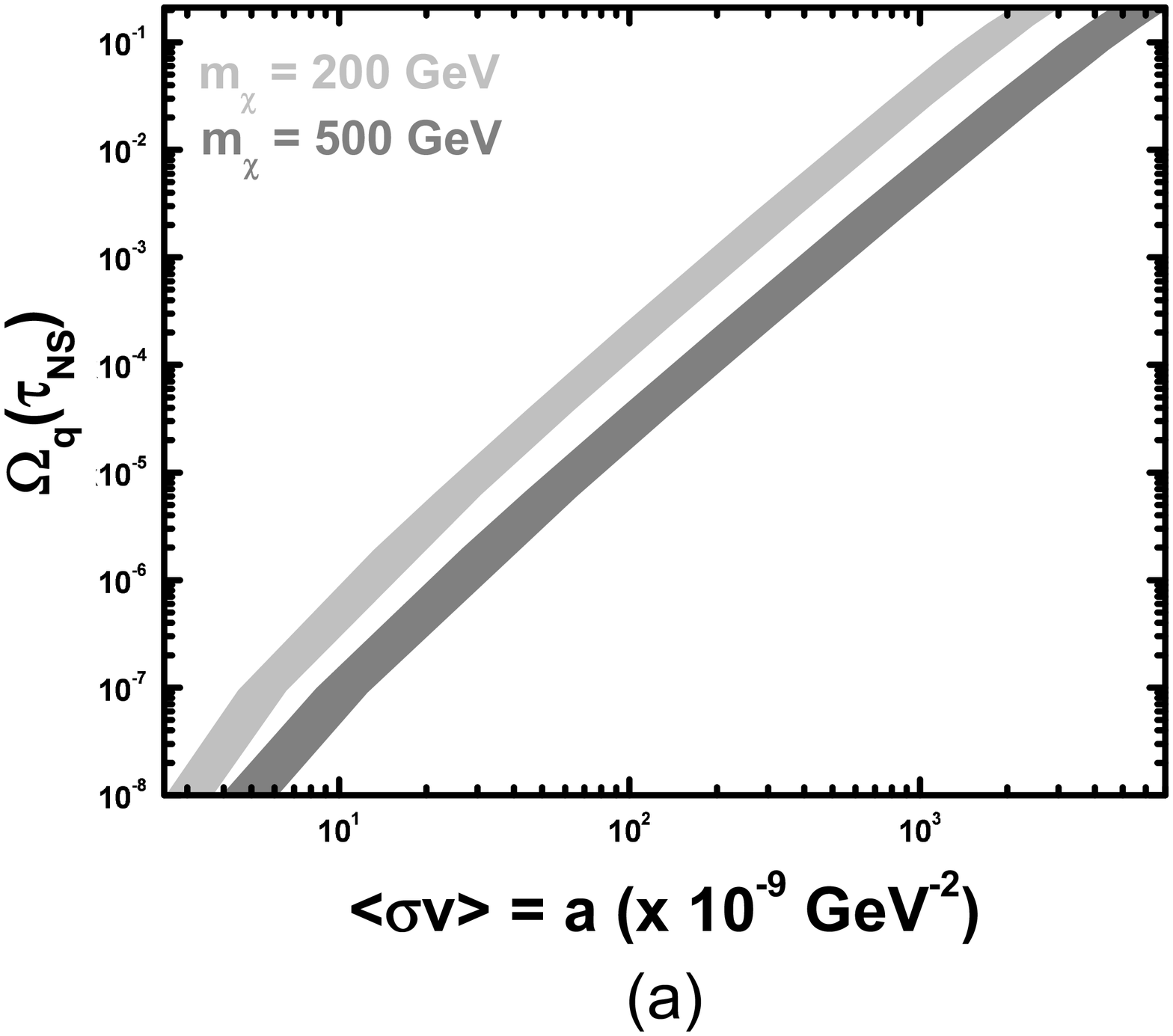,height=3.65in,angle=-90} \hspace*{-1.37 cm}
\epsfig{file=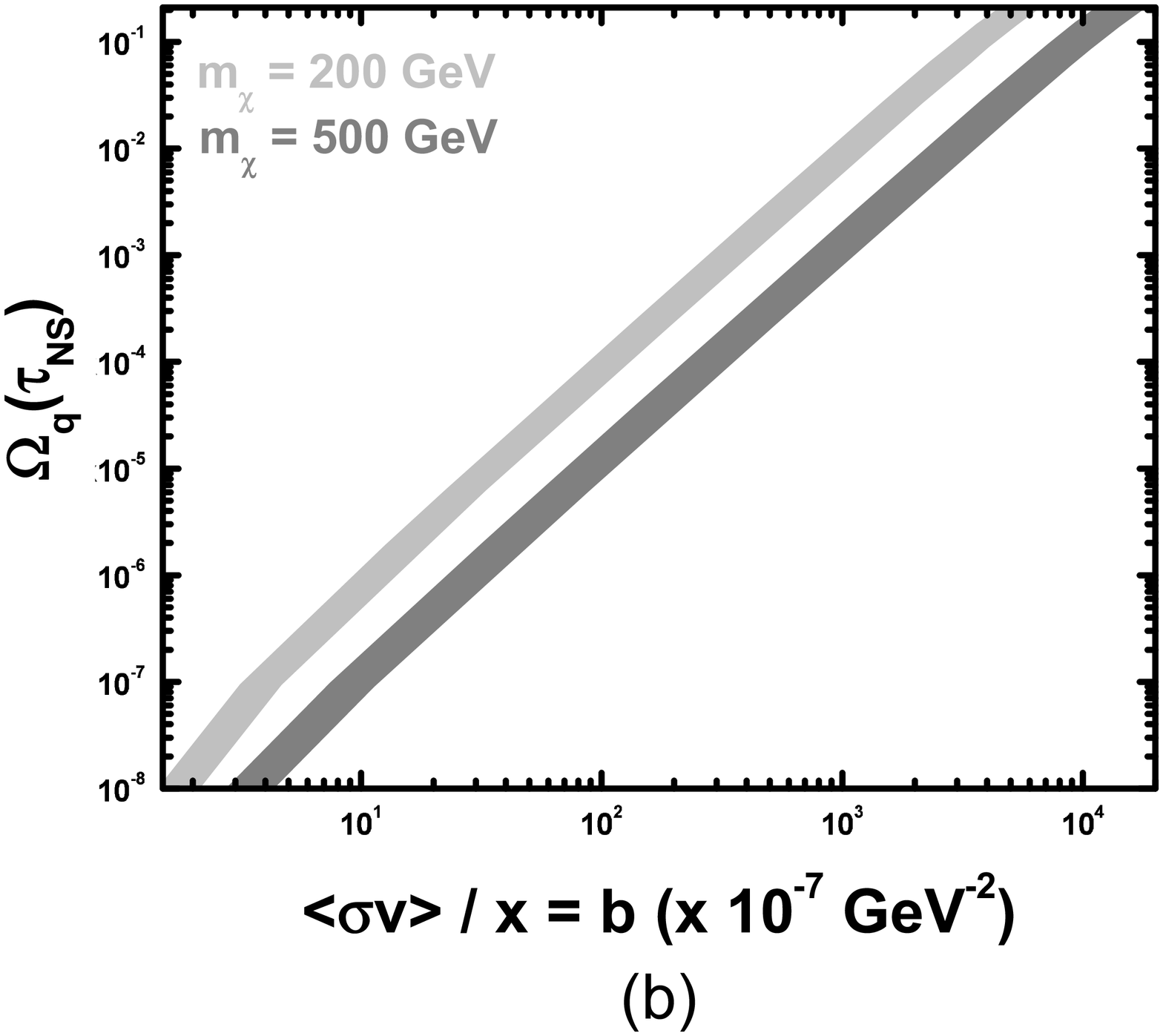,height=3.65in,angle=-90} \hfill
\end{minipage}
\hfill \caption[]{\sl The allowed regions for $m_\chi=200~{\rm
GeV}~[m_\chi=500~{\rm GeV}]$ (light [normal] grey areas) on the
$\sv-\Omega_q(\vtauf_{_{\rm NS}})$ plane for $\sv=a$ {\sf (a)} or
on the $\sv/x-\Omega_q(\vtauf_{_{\rm NS}})$ plane for $\sv=bx$
{\sf (b)}.}\label{omsv}
\end{figure}

\subsubsection{Constraining further the
$q$ parameters.}\label{qparf} Fixing the CDM parameters to
naturally obtainable values, $m_\chi=350~{\rm GeV}$ and
$\sv=10^{-7}~{\rm GeV}^{-2}$ (which yield $\Omega_\chi
h_0^2|_{_{\rm SC }}\simeq0.0025$), we can constrain further the
$q$ parameters, which are already constrained by the
quintessential requirements in subsec.~\ref{Qpar}. The regions
consistent with the achievement of eq.~(\ref{cdmb}) are ruled in
fig.~\ref{ltH}. As expected from the argument of
subsec.~\ref{omqdep}, $\Omega_q(\vtns)$ turns out to be constant
and equal to $8.5\times10^{-5}~[3.6\times10^{-5}]$ along the right
[left] boundaries of the ruled areas in fig.~\ref{ltH}-{\sf (a)},
fig.~\ref{ltH}-{\sf (b)} and along the inclined upper [lower]
boundary of the ruled area in fig.~\ref{ltH}-{\sf (c)}. If we had
imposed only the bound from eq.~(\ref{cdmb}{\sf b}),  we would
have obtained obviously much wider allowed regions bounded from
the upper boundary of ruled area and the lower boundary of the
light shaded area.

\subsubsection{Constraining further $\Omega_q(\vtns)$.}\label{omq}

Since $\Omega_\chi h_0^2$ depends exclusively on
$\Omega_q(\vtns)$'s for fixed $m_{\chi}$ and $\sv$, it would be
interesting to delineate the allowed parameter space on the
$\sv-\Omega_q(\vtns)$~[$\sv/x-\Omega_q(\vtns)$] plane for $\sv=a$
[$\sv=bx$] and fixed $m_{\chi}$. This aim is realized in
fig.~\ref{omsv}-${\sf (a)}$~[fig.~\ref{omsv}-${\sf (b)}$]. The
light [normal] grey regions are constructed for $m_{\chi}=200~{\rm
GeV}~[m_{\chi}=500~{\rm GeV}]$. Lower $m_\chi$'s require lower
$\sv$'s, since $\Delta\Omega_\chi$ decreases with $m_\chi$ as we
explain in sec.~\ref{numan}. Also, the upper [lower] boundaries of
the allowed areas are derived from eq.~(\ref{cdmb}${\sf a}$)
[eq.~(\ref{cdmb}${\sf b}$)], for the reason already mentioned in
sec.~\ref{svmp}. Consequently, when the CDM parameters are given,
restrictions on $\Omega_q(\vtns)$ supplementary to those from
eq.~(\ref{nuc}) can be derived from eq.~(\ref{cdmb}).

\section{C{\ftn ONCLUSIONS}-O{\ftn PEN} I{\ftn SSUES}}
\label{con}

\hspace{.67cm} We studied the cosmological evolution of a scalar
field $q$ which rolls down its exponential potential ensuring an
early KD epoch and acting as quintessence today. We then
investigated the decoupling of a CDM candidate, $\chi$, during the
KD epoch and calculated $\Omega_{\chi}h_0^2$. We solved the
problem (i) numerically, integrating the differential equations
which govern the cosmological evolution of $q$ and the
$\chi$-number density (ii) semi-analytically, producing
approximate relations for the former quantities. The second way
facilitates the understanding of the problem and gives, in all
cases, accurate results.

The parameters of the quintessential model ($\lambda, \vti, \vHi$)
were confined so as $0.5\leq\Omega_q(\vti)\leq1$ and were
constrained by using current observational data originating from
nucleosynthesis, the acceleration of the universe and the DE
density parameter. We found $0<\lambda<1.15$ and that there are
reasonably allowed regions on the ($\vti, \vHi$)-parameter space.
We also showed that $\Omega_{\chi}h_0^2$ increases w.r.t its value
in the SC with fixed $m_\chi$ and $\sv$. We analyzed the variation
of this enhancement, $\Delta\Omega_\chi$, w.r.t ($\lambda, \vti,
\vHi$) demonstrating that it can be expressed  as a function of
$\Omega_q(\vtns)$. We, also, found that $\Delta\Omega_\chi$
increases with $\Omega_q(\vtns)$ and $m_\chi$ and as $\sv$
decreases. It is, also, larger in the $\sv=bx$ case than in the
$\sv=a$ case for $a=b$ and fixed $m_\chi$ and $\Omega_q(\vtns)$.
By enforcing the CDM constraint, $\Omega_q(\vtns)$ close to its
upper bound requires almost three orders of magnitude larger
$\langle\sigma v\rangle$'s than those required in the SC for fixed
$m_\chi$.

Our formalism could become applicable to other more elaborated
quintessential models \cite{brax, rosati, masiero}. Also, it could
be easily extended, in order to include coannihilations and/or
$A$-pole effects for the $\Omega_{\chi}h_0^2$ calculation in the
context of specific SUSY or ED models. In the latter case, novel
deviations \cite{extra} from the SC arise, which could be
similarly analyzed (although the brane-tension is to be rather low
in order numerically visible changes on the $\Omega_{\chi}h_0^2$
calculation to be observable \cite{seto}). Also, low-reheating
scenaria \cite{fornengo, snr, kohri} could become extremely
appealing in the presence of quintessence, since they succeed to
reduce $\Omega_{\chi}h_0^2$ without need of tuning the
particle-model parameters (their coexistence with the
quintessential evolution deserves certainly deeper investigation
\cite{liddle}). On the other hand, the $\Omega_{\chi}h_0^2$
enhancement is welcome for wino or higgsino LSPs \cite{wells, su5,
nelson}, which yield $\Omega_{\chi}h_0^2$ lower than the
expectations in the SC, and so, $\Delta\Omega_\chi$ can drive
$\Omega_{\chi}h_0^2$ to the correct value \cite{prof}. In the same
time, relatively high direct detection rates can be produced
without invoking the questionable normalization \cite{nelson} of
the proton-nucleus cross section.


\ack \hspace{.67cm} The author would like to thank K. Dimopoulos,
U. Fran\c{c}a and G. Lazarides for enlightening communications,
the Greek State Scholarship Foundation (I. K. Y.) and the European
Network ENTApP under contract RII-CT-2004-506222 for financial
support.

\section*{R\footnotesize EFERENCES}

\rhead[\fancyplain{}{ \bf \thepage}]{\fancyplain{}{Q{\ftn
UINTESSENTIAL} K{\ftn INATION AND} CDM A{\ftn BUNDANCE} }}
\lhead[\fancyplain{}{R{\ftn EFERENCES} }]{\fancyplain{}{\bf
\thepage}} \cfoot{}

\end{document}